\documentclass[12pt]{iopart}
\usepackage{iopams}
\usepackage{graphicx}
\usepackage{gensymb}

\newcommand{\figref}[1]{\figurename~\ref{#1}}

\begin{document}

\title[A survey of EBW H\&{}CD potential for spherical tokamaks]{A survey of electron Bernstein wave heating and current drive potential for spherical tokamaks}
\author{ Jakub Urban$^{1}$,  Joan Decker$^{2}$, Yves Peysson$^{2}$, Josef Preinhaelter$^{1}$, Vladimir Shevchenko$^{3}$, Gary Taylor$^{4}$, Linda Vahala$^{5}$ and George Vahala$^{6}$}
\address{$^{1}$ Institute of Plasma Physics AS CR, v.v.i., Association EURATOM/IPP.CR, Prague, Czech Republic}
\address{$^{2}$ CEA, IRFM, F-13108 Saint Paul Lez Durance, France}
\address{$^{3}$ EURATOM/CCFE Fusion Association, Culham Science Centre, Abingdon, OX14 3DB, UK}
\address{$^{4}$ Princeton Plasma Physics Laboratory, Princeton, NJ 08543, USA}
\address{$^{5}$ Old Dominion University, Norfolk, VA 23529, USA}
\address{$^{6}$ College of William {\&} Mary, Williamsburg, VA 23185, USA}
\ead{urban@ipp.cas.cz}


\begin{abstract}

The electron Bernstein wave (EBW) is typically the only wave in the electron 
cyclotron (EC) range that can be applied in spherical tokamaks for heating 
and current drive (H{\&}CD). Spherical tokamaks (STs)
operate generally in 
high-$\beta $ regimes, in which the usual EC O- and X- modes are cut-off. In 
this case, EBWs seem to be the only option that can provide features similar 
to the EC waves---controllable localized H{\&}CD that can be utilized for 
core plasma heating as well as for accurate plasma stabilization.

The EBW is a quasi-electrostatic wave that can be excited by mode conversion 
from a suitably launched O- or X-mode; its propagation further inside the 
plasma is strongly influenced by the plasma parameters. These rather awkward 
properties make its application somewhat more difficult. In this paper we 
perform an extensive numerical study of EBW H{\&}CD performance in four 
typical ST plasmas (NSTX L- and H-mode, MAST Upgrade, NHTX). Coupled 
ray-tracing (AMR) and Fokker-Planck (LUKE) codes are employed to simulate 
EBWs of varying frequencies and launch conditions, which are the fundamental 
EBW parameters that can be chosen and controlled. Our results indicate that 
an efficient and universal EBW H{\&}CD system is indeed viable. In 
particular, power can be deposited and current reasonably efficiently driven 
across the whole plasma radius. Such a system could be controlled by a 
suitably chosen launching antenna vertical position and would also be 
sufficiently robust.

\end{abstract}
\submitto{Nuclear Fusion}

\section{Introduction}
\label{sec:introduction}
Present research in electron cyclotron (EC) wave heating and current drive 
(H{\&}CD) for magnetic confinement thermonuclear fusion \cite{1} is focused on 
conventional aspect ratio tokamaks, and particularly ITER. However, in the 
``alternative'' spherical tokamak (ST) with aspect ratio $A\equiv R_0 /a$ 
close to unity ($R_0 $ and $a$ being the major and minor radii, 
respectively) and weaker external toroidal magnetic field, the usual EC 
transverse O- and X-modes are mostly cut-off and cannot be used for H{\&}CD. 
The role of ST research is nevertheless very important. For their relatively 
high neutron flux density and economy, STs are being considered as a 
candidate for a component test facility (ST-CTF) \cite{2, 3} and, for the same reasons, 
appear in fusion-fission hybrid concepts \cite{4}.

ST's low magnetic field has a major impact on the propagation of electron 
cyclotron waves in the plasma. This frequency range is of crucial importance for 
auxiliary H{\&}CD systems in present and future tokamaks. Typically, in STs, 
the electron plasma frequency ${\omega _{{\mathrm{pe}}}} = 2\pi {f_{{\mathrm{pe}}}} \equiv {\left( { {n_{\mathrm{e}}}{e^2}/{m_{\mathrm{e}}\varepsilon_0}} \right)^{1/2}}$ is much 
greater than the electron cyclotron frequency ${\omega _{{\mathrm{ce}}}} = 2\pi {f_{{\mathrm{ce}}}} \equiv eB/{m_{\mathrm{e}}}$. Here, $n_\mathrm{e} $ is the electron 
density, $B$ is the total magnetic field, $e$ is the electron charge, 
$m_\mathrm{e} $ is the electron mass, and $\varepsilon_0$ is the vacuum permittivity. A 
similar situation often arises in stellarators, which do not have any 
principal MHD stability density limits. In this so called overdense regime, 
particularly when $\omega _{\mathrm{pe}} >n\omega _{\mathrm{ce}} $, $n>3$ in 
most of the plasma cross section, the O- and X-modes of EC waves with 
1$^\mathrm{st}$, 2$^\mathrm{nd}$ and 3$^\mathrm{rd}$ harmonic frequencies ($\omega _{\mathrm{ce}} 
<\omega <4\omega _{\mathrm{ce}} )$ are cut-off and cannot propagate inside the 
overdense plasma. Higher harmonic EC waves are not of interest because of 
their very weak absorption (low optical depth). However, the electron Bernstein wave (EBW) \cite{5}---a 
quasi-electrostatic kinetic EC mode---can propagate and be strongly absorbed 
in an overdense plasma.

EC waves are extremely useful because they can be launched far from the 
plasma (they do not need large plasma-facing antenna structures like 
ion-cyclotron or lower-hybrid waves) and feature highly localized and 
controllable H{\&}CD capabilities. The application of the overdense 
mode---the EBW---is, however, complicated by its electrostatic nature. 
First, EBWs must be excited by appropriately launched O- or X-mode via so 
called OXB or XB mode conversion scheme. This mode conversion takes place in 
the upper hybrid resonance region, where the wave frequency satisfies 
$\omega =\omega _{\mathrm{UH}} \equiv \sqrt {\omega _{\mathrm{pe}}^2 +\omega 
_{\mathrm{ce}}^2 } $. This typically occurs near the plasma edge. The mode 
conversion efficiency depends on the wave and plasma parameters and is thus 
a potential source of power loss. The excited EBW can subsequently propagate 
inside the overdense plasma; however, because of its dispersion 
characteristics, the propagation strongly depends on plasma parameters and 
the wave vector can change considerably in various ways (unlike O- and 
X-mode propagation, during which the parallel wave number is mostly 
conserved).

H{\&}CD by EBWs have already been demonstrated experimentally in magnetic 
confinement fusion devices, particularly in COMPASS-D \cite{6} and Wendelstein 
7-AS \cite{7, 8}.
Numerical studies of advanced (steady-state) spherical tokamak 
operations consider EBWs as one of the vital current drive system that can 
stabilize MHD instabilities \cite{9}.

In this paper, we pursue an overall study of EBW H{\&}CD on spherical 
tokamaks by means of numerical ray-tracing and Fokker-Planck simulations. 
Two coupled codes---AMR (Antenna, Mode-conversion, Ray-tracing) \cite{10, 11} and 
LUKE \cite{12, 13}---are employed. These codes have proven to be very suitable 
for EBWs \cite{14, 11}. A large number of cases with different injection 
parameters is simulated in four different ST conditions: two experimental 
discharges of the NSTX tokamak \cite{15}, an ST-CTF-like MAST-Upgrade H-mode scenario \cite{16,16b} and 
an NHTX scenario \cite{15}. The resulting extensive collection of results is 
analyzed with emphasis on H{\&}CD performance---viability, effectiveness, 
flexibility, controllability and robustness. 

The paper is organized into five sections. After the introduction, the 
description of EBW physics involved in this study is given. Section 
\ref{sec:simulated} describes the context of the numerical 
simulation---the target plasmas and the EBW parameters. Numerical 
results---particularly the current drive localization and efficiency, the 
role of $N_\parallel $, $Z_{\mathrm{eff}} $, the quasilinear effects and the 
robustness---are presented in section \ref{sec:mylabel1}. Finally, 
section \ref{sec:conclusions} discusses the conclusions of our work.

\section{Electron Bernstein wave description}
\label{sec:electron}
The electron Bernstein wave (EBW) has been known for decades \cite{5}. It appears 
as another EC branch, besides the cold plasma O- and X-modes, when solving 
the kinetic dispersion relation of a plasma in an external magnetic field. 
It was later applied to magnetic confinement fusion for H{\&}CD while EBW 
emission was introduced as a diagnostic tool; an overview of EBW experiments 
is given in \cite{17}. EBW physics can basically be separated into three areas: 
the mode conversion, the propagation and the wave-plasma power transfer.

\subsection{Mode-conversion}
\label{subsec:mylabel1}
The EBW mode conversion is a full wave process in which transverse electron 
cyclotron modes and the EBW are involved. The mode conversion always 
requires the slow X-mode to be excited, which can then fully convert into 
the EBW at the upper hybrid resonance (UHR). This study is confined to EBW 
excitation from the low-field side (LFS) of a tokamak, where two 
possibilities exist: the XB \cite{18} and the OXB \cite{19, 20} conversion. The XB 
conversion is characterized by a direct coupling between the fast and the 
slow X-mode while in the OXB scheme the O-mode is converted to the slow 
X-mode (in fact, it is converted to the fast branch of the X-mode, which 
propagates towards higher density and then smoothly converts to the slow 
branch, which propagates backwards to the UHR). The XB scheme is typically 
efficient for lower frequencies and requires the density scale length to be 
specifically adjusted (for details see \cite{18}). The OXB scheme, on the other 
hand, is more universal in terms of frequencies and density scale lengths as 
efficient conversion only requires the O-mode to be incident at the optimum 
angle. In 1D theory, this angle is given by \cite{19, 21}
\begin{equation}
\label{eq1}
N_{\parallel \mathrm{opt}}^2 =\left( {1+\omega /\omega _{\mathrm{ce}} } 
\right)^{-1},\;N_{\mathrm{pol}} =0,
\end{equation}
where ${\rm {\bf N}}\equiv c{\rm {\bf k}}/\omega $ is the normalized wave 
vector, $N_\parallel \equiv {\rm {\bf N}}\cdot {\rm {\bf B}}/B$, and 
$N_{\mathrm{pol}} \equiv {\rm {\bf N}}\cdot \left( {{\rm {\bf B}}\times \nabla 
n_\mathrm{e} } \right)/\left\| {{\rm {\bf B}}\times \nabla n_\mathrm{e} } 
\right\|$. If the incident angle is not optimum, the O- to X-mode power 
conversion efficiency of a plane wave decays approximately exponentially as 
\cite{20, 21}
\begin{equation}
\label{eq2}
C_{\mathrm{OX}} =e^{-\pi k_0 L_n \sqrt {\omega _{\mathrm{ce}} /2\omega } \left( 
{2\left( {1+\omega _{\mathrm{ce}} /\omega } \right)\left( {N_{\parallel 
\mathrm{opt}} -N_\parallel } \right)^2+N_{\mathrm{pol}}^2 } \right)}
\end{equation}
where $L_n \equiv n_\mathrm{e} /\left( {\mathrm{d}n_\mathrm{e} /\mathrm{d}r} 
\right)$ is the density scale length, and (\ref{eq2}) is evaluated at the O-mode 
cut-off. 
This formula agrees well with numerical results for $k_0 L_n>1$ when 
$N_\parallel$ is almost optimum \cite{20a} so that $C_{\mathrm{OX}}>0.9$.
When $N_\parallel$ is further from the optimum, larger $k_0 L_n$ is required
for good agreement \cite{20a}.
A 1D formalism has also been developed \cite{21a} which solves the mode-conversion
problem for almost arbitrary inhomogeneity scale length.
In general conditions, the conversion efficiency must be calculated 
by full-wave codes, either in the cold plasma \cite{22} or the hot plasma \cite{23} 
approximation. These codes also take into account the incident wave 
polarization and the parasitic slow to fast X-mode tunneling, which can 
decrease the OXB conversion efficiency, especially when the density gradient
is very steep and consequently $k_0 L_n$ is small. 
Numerical and analytical results are 
in good agreement particularly around the optimum incidence. As a 
consequence of the exponential efficiency dependence, there always exists an 
angular window where the mode conversion is sufficiently effective.

Recently, 2D theory and simulations of the OXB conversion have been 
developed \cite{24,25,26,27}. 2D effects are shown to be important for off-equatorial 
launch, where the O-mode cutoff and X-mode cutoff surfaces are no longer 
parallel \cite{26}. In \cite{27}, it is shown that the beam curvature should be 
matched to the plasma surface curvature in order to optimize the conversion 
efficiency. The effect of the beam size, which obviously determines the beam 
spectrum, is also studied, showing that larger beams, i.e. narrower 
$k$-spectrum, tend to be more efficiently converted \cite{27}. Non-linear effects 
can also play a role, for example a parametric decay \cite{28} or higher harmonic 
wave generation \cite{29}. Another factor typically not considered in the 
conversion process is density fluctuations. An estimate, based on a 
probability distribution function and the 1D formula (\ref{eq2}), was made and 
experimentally demonstrated in \cite{7}, showing a considerable decrease of the 
conversion efficiency for large density scale lengths. A detailed treatment 
of this problem should employ a 2D approach.

In this paper, even though the mode conversion is not treated in detail, it 
is nonetheless not neglected. We consider the OXB scheme for its 
universality. However, the results are directly applicable to any
mode conversion scheme as long as EBWs are excited at the same place
with similar $N_\parallel$.
The OXB scheme was successfully demonstrated in various past and 
present experiments \cite{17}. Our EBW H{\&}CD simulation starts from an antenna, 
which emits a Gaussian beam \cite{30} of a given frequency and waist radius $w_0 
$, which, in our case, is calculated from the Rayleigh range 
\begin{equation}
\label{eq3}
z_\mathrm{R} \equiv \pi w_0^2 /\lambda _0 =k_0 w_0^2 /2,
\end{equation}
where $\lambda _0 $ and $k_0 $ are the vacuum wavelength and the wavenumber, 
respectively. At the distance $z_\mathrm{R} $ from the waist, the Gaussian 
beam doubles its spot size (the beam radius becomes $\sqrt 2 w_0 )$. At a 
fixed $z_\mathrm{R} $, the beam divergence is similar for all frequencies. 
Large divergence, i.e., wide beam $k$-spectrum, would cause poor O-X 
conversion efficiency.

We now calculate the conversion efficiency of a Gaussian beam. The electric 
field of a Gaussian beam in the Fourier space is
\begin{equation}
\label{eq4}
E\propto \exp \left( {-\frac{w_0^2 }{4}\left( {k_x^2 +k_y^2 } \right)} 
\right),
\end{equation}
where $k_{x,y} $ are wave vectors perpendicular to the direction of the beam 
propagation. The corresponding energy density is $\propto \left| E 
\right|^2$. We can evaluate the total power of the O-X converted Gaussian 
beam using the analytic formula (\ref{eq2}) and Parseval's theorem:
\begin{equation}
\label{eq5}
\frac{P_{\mathrm{OX}} }{P_0 }=\frac{\int_{-\infty }^\infty {\int_{-\infty 
}^\infty {\left| E \right|^2C_{\mathrm{OX}} \mathrm{d}k_x \mathrm{d}k_y } } 
}{\int_{-\infty }^\infty {\int_{-\infty }^\infty {\left| E 
\right|^2\mathrm{d}k_x \mathrm{d}k_y } } }.
\end{equation}
We assume here that the error introduced by changing the integration limits 
from $k_x^2 +k_y^2 \leqslant k_0^2 $ to $\pm \infty $ is negligible, which 
is valid for beams with not too large divergence. We now assume that the ${\rm {\bf 
\hat {z}}}$ axis is along the beam propagation direction and that the launch 
angle is optimum by setting, without any loss of generality, the magnetic field at the 
O-mode cut-off to be ${\rm {\bf B}}/B=\left( {0,\pm \sqrt {1-N_{\parallel 
,\mathrm{opt}}^2 } ,N_{\parallel ,\mathrm{opt}} } \right)$.
I.e., the magnetic field is assumed to be homogeneous across the beam spot at the O-mode cutoff.
The integral (\ref{eq5}) then becomes
\begin{equation}
\label{eq6}
\fl
\frac{P_{\mathrm{OX}} }{P_0 }=\frac{\int_{-\infty }^\infty {\int_{-\infty 
}^\infty {\left| E \right|^2e^{-\frac{\pi L_n \sqrt {\omega _{\mathrm{ce}} } 
}{k_0 \sqrt {2\omega } }\left( {2\left( {1+\omega _{\mathrm{ce}} /\omega } 
\right)\left( {k_0 N_{\parallel ,\mathrm{opt}} \mp k_y \sqrt {1-N_{\parallel 
,\mathrm{opt}}^2 } -\sqrt {k_0^2 -k_x^2 -k_y^2 } N_{\parallel ,\mathrm{opt}} } 
\right)^2+k_x^2 } \right)}\mathrm{d}k_x \mathrm{d}k_y } } }{\int_{-\infty 
}^\infty {\int_{-\infty }^\infty {\left| E \right|^2\mathrm{d}k_x \mathrm{d}k_y 
} } }.
\end{equation}
This can be evaluated analytically after Taylor-expanding the exponent in 
$k_{x,y} $ around 0 to second order, yielding
\begin{equation}
\label{eq7}
\frac{P_{\mathrm{OX}} }{P_0 }=\left( {1+3\frac{\kappa }{z_\mathrm{R} 
}+2\frac{\kappa ^2}{z_\mathrm{R}^2 }} \right)^{-\frac{1}{2}}
\end{equation}
where $\kappa \equiv \pi L_n \sqrt {\omega _{\mathrm{ce}} /2\omega } $. This 
is an important result which, in fact, imposes an upper limit to the 
conversion efficiency of a Gaussian beam. This limit depends on $L_n 
/z_\mathrm{R} $ for a fixed $\omega _{\mathrm{ce}} /\omega $. (Note that 
$0.4<\sqrt {\omega _{\mathrm{ce}} /2\omega } <0.7$ for the first two 
harmonics). It also tells us how narrow (i.e. how divergent) a beam can be used 
while keeping the OXB conversion efficient. The beam conversion efficiency 
is shown graphically for the discussed scenarios in section 
3.2. In a similar fashion, we can also evaluate the 
conversion efficiency of a Gaussian beam for non-optimum central wave 
vector. 

\subsection{Propagation}
\label{subsec:propagation}
The electron Bernstein wave is, apart from the ``cold'' O- and X-modes, a 
solution to the kinetic (hot) dispersion relation of a plasma in an external 
magnetic field \cite{5}. Numerous analytical and numerical studies of EBW 
propagation have been performed and EBW propagation is hence quite well 
explored. The characteristic properties of EBW propagation are:

\begin{itemize}
\item The polarization is quasi-electrostatic; hence, in most situations, the electrostatic dispersion relation describes EBWs satisfactorily \cite{31, 32, 32a}.
\item EBWs can propagate in plasmas if $\omega _{\mathrm{ce}} <\omega \lesssim \omega _{\mathrm{UH}} $\footnote{$\omega < \omega _{\mathrm{UH}}$ holds for $\omega < 2\omega_\mathrm{ce}$. Higher frequency EBWs can propagate in a rarified plasma up to a certain lower electron density limit.}. As $\omega \sim n\omega _{\mathrm{ce}} $, $n=1,2,\ldots $, it implies $n\omega _{\mathrm{ce}} <\omega _{\mathrm{UH}} $. Note again that EC O- and X-mode are mostly cut-off under these conditions. There is no upper density cut-off for EBWs.
\item The phase velocity is almost perpendicular to the external magnetic field ($k_\bot \gg k_\parallel )$; the group velocity is generally different from the phase velocity in both magnitude and direction.
\item The wavelength is of the order of the electron gyroradius $\rho _\mathrm{e} \equiv v_{\mathrm{Te}} /\omega _{\mathrm{ce}} $ ($v_{\mathrm{Te}} \equiv \sqrt {T_\mathrm{e} /m_\mathrm{e} } $ is the electron thermal velocity and $T_\mathrm{e} $ is the electron temperature in energy units), i.e.,
\end{itemize}
\begin{equation}
\label{eq8}
k\rho _\mathrm{e} \cong k_\bot \rho _\mathrm{e} \sim 1.
\end{equation}
\begin{itemize}
\item EBW characteristics vary significantly depends on whether it approaches a resonance at the lower harmonic ($\omega >n\omega _{\mathrm{ce}} )$ or at the higher harmonic ($\omega <n\omega _{\mathrm{ce}} $). In particular, the perpendicular wave vector is much smaller near the resonance if $\omega <n\omega _{\mathrm{ce}} $, where $k_\bot \rho _\mathrm{e} \ll 1$ and electromagnetic effects are no longer negligible. Yet, as the wave approaches the resonance, the power is usually absorbed before the electrostatic approximation becomes invalid \cite{31}.
\item The parallel refractive index $N_\parallel $ evolves during the propagation and can be greater than one (unlike O- and X-modes). Depending on the magnetic field topology and the vertical launch position, the wave parallel index can either stay close to its initial value, or oscillate around zero, or increase/decrease steadily. During the actual EBW propagation, the wave parallel index can change the regime. In general, waves close to the midplane tend to have a flat or oscillating $N_\parallel $ \cite{33}, while for off-midplane rays $N_\parallel $ increases or decreases steadily at a rate proportional to the distance from the midplane \cite{18}. This property forms the basis for controlling EBWs.
\end{itemize}
EBW propagation in a tokamak plasma is far from trivial and necessitates a 
numerical simulation. The ray-tracing technique is well suited for EBW 
propagation since the WKB validity conditions are well fulfilled due to the 
short wavelength. We employ the AMR code \cite{10, 11} to simulate the EBW 
propagation. This code uses a conventional ray-tracing method \cite{34, 35} with 
an electrostatic kinetic non-relativistic dispersion relation \cite{36}:
\begin{equation}
\label{eq9}
\mathcal{D}\equiv 1+\left( {\frac{\omega _{\mathrm{pe}}^2 }{k^2v_{\mathrm{Te}}^2 
}} \right)\left( {1+\sum\limits_n {\frac{\omega }{\sqrt 2 k_\parallel 
v_{\mathrm{Te}} }e^{-b}I_n \left( b \right)Z\left( {\xi _n } \right)} } 
\right)=0,
\end{equation}
where
\begin{equation}
\label{eq10}
\xi _n \equiv \frac{\omega +n\omega _{\mathrm{ce}} }{\sqrt 2 \left| 
{k_\parallel } \right|v_{\mathrm{Te}} },\quad b\equiv \left( {\frac{k_\bot 
v_{\mathrm{Te}} }{\omega _{\mathrm{ce}} }} \right)^2,
\end{equation}
$Z$ is the plasma dispersion function \cite{37} and $I_n $ is the modified Bessel 
function of the first kind. The ray trajectory is then a solution to the 
Hamiltonian-type equations
\begin{equation}
\label{eq11}
\eqalign{
\frac{\mathrm{d}{\rm {\bf r}}}{\mathrm{d}t} & = -\frac{\partial \Re{}\left( \mathcal{D} \right)}{\partial {\rm {\bf k}}}/\frac{\partial \Re{}\left( \mathcal{D} \right)}{\partial \omega },\\
\frac{\mathrm{d}{\rm {\bf k}}}{\mathrm{d}t} & = \frac{\partial \Re \left( \mathcal{D} \right)}{\partial {\rm {\bf r}}}/\frac{\partial \Re{}\left( \mathcal{D} \right)}{\partial \omega }.
}
\end{equation}
Equations (\ref{eq11}) conserve $\mathcal{D}=0$ at zeroth 
order in the parameter $\Im{}\left( \mathcal{D} \right)/\Re{}\left( \mathcal{D} 
\right)$, i.e., in a weak damping approximation. 

In this study, we neglect the collisional damping, which can, however, be of 
critical importance, as was previously shown by our modelling and by 
experiments at NSTX \cite{38, 39}. Collisional damping is extremely sensitive to 
edge plasma conditions, particularly the temperature (or, more precisely, 
the collision frequency) in the mode conversion layer. As these conditions 
cannot be sufficiently accurately predicted, we completely ignore the effect 
of collisions.

By using the non-relativistic electrostatic dispersion (\ref{eq9}), we neglect 
relativistic and transverse electric field effects in the wave propagation. 
These effects were studied in \cite{40, 31}, and also in \cite{41}, where a 
fully-relativistic electromagnetic ray-tracing code was used. Some 
differences are seen in the ray propagation and in the polarization. These 
differences are, however, rather small and would not change the overall 
picture of our survey. The fully-relativistic approach is also 
computationally very intensive and would require an extensive amount of 
computation time to simulate all the cases presented here.

\subsection{Wave absorption}
\label{subsec:mylabel2}
In this section, we describe the theory of the EBW absorption. EBWs can be 
absorbed by resonant electrons, which satisfy the resonance condition 
\begin{equation}
\label{eq12}
\omega -n\omega _{\mathrm{ce}} /\gamma -k_\parallel v_\parallel =0,
\end{equation}
where $\gamma \equiv \left( {1-v^2/c^2} \right)^{-1/2}$ is the usual 
relativistic factor. While the absorption mechanism---the EC harmonic 
damping---is identical to that for O- and X-modes, the polarization of EBWs 
is different (quasi-electrostatic), and this leads to a strong interaction 
even for low temperatures at any EC harmonic. In the ray-tracing approach, 
the wave absorption is assumed to be weak and can thus be handled 
independently. This requires the anti-Hermitian part of the dielectric 
tensor to be much smaller than the Hermitian part, or, alternatively, 
$\Im{}\left( \mathcal{D} \right)\ll \Re{}\left( \mathcal{D} \right)$.

The zeroth order solution of the dispersion relation leads, as shown in the 
previous section, to the ray-tracing equations (\ref{eq11}). The first order 
solution then leads to the radiative transfer equation \cite{34}
\begin{equation}
\label{eq13}
\frac{\mathrm{d}P}{\mathrm{d}t}=\vartheta -\alpha P,
\end{equation}
where 
\begin{equation}
\label{eq14}
\alpha \equiv -\frac{2\Im\left( \mathcal{D} \right)}{\left| {\partial 
\Re{}\left( \mathcal{D} \right)/\partial \omega } \right|}
\end{equation}
is the absorption coefficient, $\vartheta $ the emissivity and $P$ the ray power. 
The non-relativistic dispersion function (\ref{eq9}) is, however, inappropriate for 
the damping calculations. For this reason, the ray-tracing code rather 
employs the weakly-relativistic absorption coefficient by Decker and Ram 
\cite{31}, which is very fast and sufficiently accurate. Assuming high incident ray 
power $P_0 $, the emissivity can be neglected and the solution to \Eref{eq13} 
is
\begin{equation}
\label{eq15}
P=P_0 e^{-\int_0^\infty {\alpha \mathrm{d}t} }.
\end{equation}
Since equation (\ref{eq15}) provides a linear damping solution, any modifications to 
the electron distribution function by the waves are not taken into account. 
However, quasilinear effects due to the modified distribution function can play 
an important role in EC H{\&}CD. For this reason, we employ LUKE---a fully 
relativistic, bounce-averaged, 3-D Fokker-Planck solver \cite{12}---which 
calculates the evolution of the electron distribution function for 
axisymmetric plasmas in the low-collisionality regime. LUKE particularly 
accounts for collisions and quasilinear diffusion due to RF waves. The code 
uses a fully-relativistic, parallel momentum conserving collision operator
and a fully implicit 3-D time evolution scheme for a fast convergence to 
the time-asymptotic solution. It has been verified that the damping profile 
calculated by LUKE in the low power limit agrees with linear theory \cite{13}. 

\subsection{Simulation model}
\label{subsec:simulation}
Our model antenna emits a Gaussian beam, parameterized by its frequency, 
beam waist vertical and radial position $Z_\mathrm{A} $, $R_\mathrm{A} $ and 
waist radius $w_0 $. The beam radius is, in our case, calculated from the 
Rayleigh range, therefore fixing the beam divergence and consequently the 
beam O-X conversion efficiency (putting aside the variable density scale 
lengths and the $\omega _{\mathrm{ce}} /\omega $ dependence) for all 
frequencies. 

The antenna angles used in the following simulations are optimized for the 
OXB mode conversion, i.e., determined by the condition (\ref{eq1}) for each beam 
waist position. The average over a single harmonic frequency range ($n\omega 
_{\mathrm{ce}} <\omega <\left( {n+1} \right)\omega _{\mathrm{ce}} )$ is chosen 
as the optimum angle for these frequencies rather than the optimum evaluated 
for a particular frequency. The differences in the resulting ray-tracing 
initial conditions are negligible and central ray mode-conversion 
efficiencies do not drop below 90 {\%} in most cases. The launch conditions 
are also checked using AMR's 1D full-wave calculation \cite{22}.

The beam is discretized by a bundle of 16 individual rays. We denote the  
positions of the intersections of the rays with the beam waist
by ${\rm {\bf r}}_i^0 $, $i=1\ldots 16$. Using a straight 
line propagation, the intersection of the central ray with the O-mode 
cut-off is found. At this point, the Gaussian beam size is calculated and 
the beam is again discretized by the same, proportionally scaled 16 ray 
pattern. We denote these ray spot positions ${\rm {\bf r}}_i^1 $. The 
intersections of the straight lines, connecting ${\rm {\bf r}}_i^0 $ and 
${\rm {\bf r}}_i^1 $, with the O-mode cut-off surface are used as the 
starting points for the ray-tracing: i.e., we assume straight propagation of 
the O-mode from the beam waist to the O-mode cutoff surface, but still 
taking the beam divergence into account. The initial wave vectors for the 
ray-tracing are found by solving the electrostatic dispersion relation (\ref{eq9}) 
with respect to $N_\perp$, using
\begin{equation}
\label{eq16}
N_{\parallel 0} ={\rm {\bf N}}_\mathrm{0} \cdot {\rm {\bf B}}/B,
\end{equation}
where ${\rm {\bf N}}_0 =\left( {{\rm {\bf r}}_i^1 -{\rm {\bf r}}_i^0 } 
\right)/\left\| {{\rm {\bf r}}_i^1 -{\rm {\bf r}}_i^0 } \right\|$ is the ray 
vacuum wave vector and ${\rm {\bf B}}$ is evaluated at the O-mode cut-off.

After finding the initial wave vector, the electrostatic ray-tracing is 
started. The principal results are the ray trajectories and wave vector 
evolutions. When ray-tracing is finished, the outputs are passed to LUKE, 
along with the magnetic equilibrium and plasma profiles. The AMR-LUKE 
interface has been particularly verified to keep all quantities consistent. 
Besides, the interface is user friendly and LUKE can be launched by AMR, and 
vice versa, by a single option in the configuration file.

Finally, LUKE determines a distribution function $f_{\mathrm{ql}} $ that is 
consistent with the quasilinear wave absorption. Besides the flux-averaged 
absorbed power density profile $P_\mathrm{d} \left( \rho \right)$, the 
flux-averaged EBW-driven current density profile 
\begin{equation}
\label{eq17}
j_\parallel \left( \rho \right)=-e\left\langle {\int {v_\parallel 
f_{\mathrm{ql}} \mathrm{d}p^3} } \right\rangle 
\end{equation}
is also calculated. Here, $\left\langle \cdot \right\rangle $ denotes flux 
surface averaging and $\rho $ is a flux surface coordinate based on the 
poloidal magnetic flux $\psi $:
\begin{equation}
\label{eq18}
\rho =\rho _{\mathrm{pol}} \equiv \sqrt {\frac{\psi -\psi _{\mathrm{axis}} 
}{\psi _{\mathrm{LCFS}} -\psi _{\mathrm{axis}} }} ,
\end{equation}
where $\psi _{\mathrm{axis}} $ and $\psi _{\mathrm{LCFS}} $ is the poloidal 
magnetic flux at the magnetic axis and at the last closed flux surface 
(LCFS), respectively.

\section{Simulated scenarios of EBW H{\&}CD}
\label{sec:simulated}
\subsection{Fundamental target plasma parameters}
\label{subsec:fundamental}
As already mentioned in the introduction, EBW H{\&}CD is simulated here in 
four different target spherical tokamak scenarios, whose fundamental 
parameters are listed in \tablename~\ref{tab1}. As can be seen, 
the chosen scenarios differ in various fundamental parameters. Two of them 
are typical NSTX L- and H-mode experimental discharges, the other two are 
TRANSP (a plasma transport code) \cite{transp} model scenarios of the planned MAST Upgrade \cite{16b} and of NHTX (a potential 
plasma facing component test facility).

\begin{table}[htbp]
\caption{\label{tab1}Fundamental parameters of the target ST scenarios:
major radius $R_0$ [m], minor radius $a$ [m], central toroidal 
magnetic field $B_0 $ [T], central electron density $n_{\mathrm{e0}} $ [10$^{19}$ m$^{-3}$], central 
temperature $T_{\mathrm{e0}} $ [keV], plasma current $I_\mathrm{p} $ [MA].}
\begin{indented}
\lineup
\item[]\begin{tabular}{@{}*{8}{l}}
\br
Name&
$R_0$ &
$a$ & 
$B_0 $& 
$n_{\mathrm{e0}} $& 
$T_{\mathrm{e0}} $& 
$I_\mathrm{p} $& 
Origin \\
\mr
NSTX L-mode& 
1.0 &
0.52 & 
0.5& 
2.6& 
2.9& 
0.6 & 
shot 123435 \\
NSTX H-mode& 
1.0 &
0.5 & 
0.5& 
3.9& 
1.4& 
1.0& 
shot 130607 \\
MAST Upgrade& 
0.93 &
0.41 & 
0.78& 
3.5& 
2.4& 
1.2& 
TRANSP \\
NHTX& 
1.2 &
0.37 & 
2.0& 
20.0& 
5.7& 
3.5& 
TRANSP \\
\br
\end{tabular}
\end{indented}
\end{table}

\begin{figure}[htbp]
\centering
\includegraphics[width=0.9\linewidth]{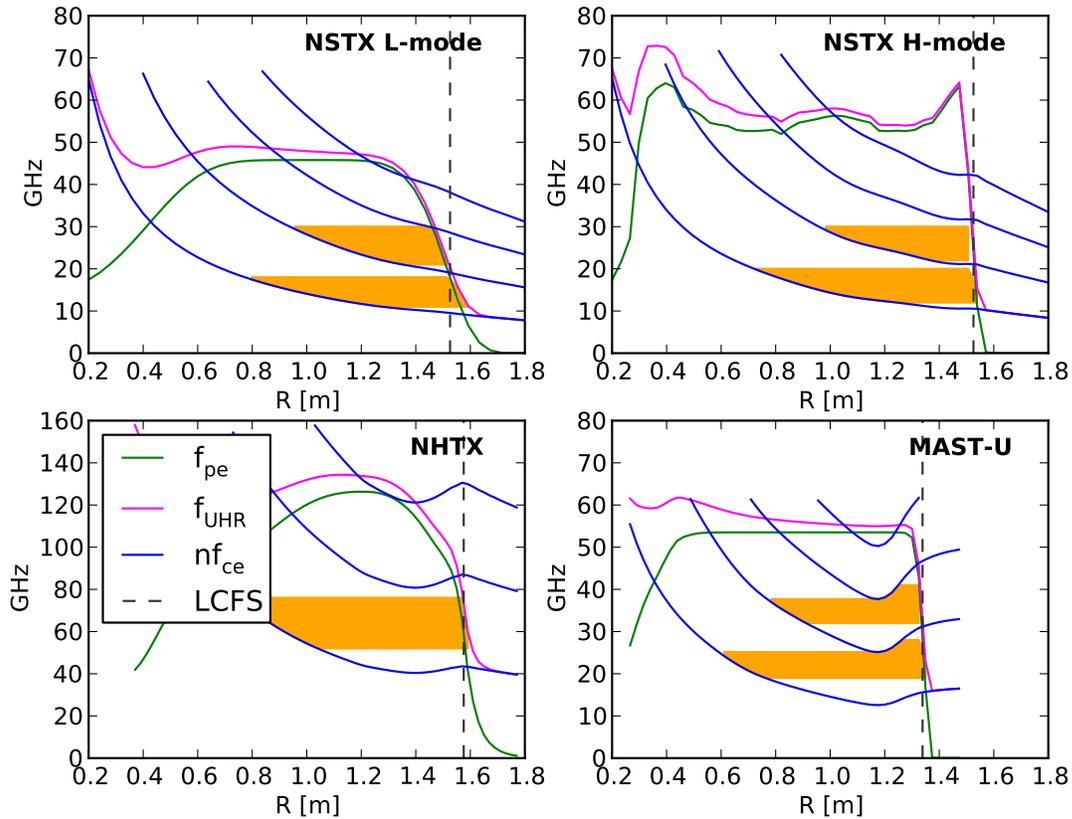}
\caption{Radial profiles (in the mid-plane) of the characteristic frequencies for the target scenarios. $f_{\mathrm{pe}}$ -- electron plasma frequency, $f_{\mathrm{UHR}} $ -- upper hybrid frequency, $nf_{\mathrm{ce}} $ -- n$^\mathrm{th}$ EC harmonic. Filled areas represent the simulated frequency ranges.}
\label{fig1}
\end{figure}

In \figref{fig1} are plotted the midplane radial 
profiles of the characteristic frequencies for the target scenarios. The 
simulated frequency ranges are marked by shaded areas, whose 
areas delimit the EBW propagation regions, i.e., from the UHR at the edge
to the cold EC resonance, which is theoretically accessible by $N_\parallel=0$
waves only.
Clearly, the plasma is overdense ($f_{\rm pe} > nf_{\rm ce}$) and the first three EC harmonics are 
inaccessible for O- and X-modes in NSTX and MAST. In NHTX, the third 
harmonic is more or less accessible and the corresponding frequency is 
compatible with present-day gyrotron technology. The first two EC harmonics 
have been selected for NSTX and MAST, as higher harmonics will likely be 
overlapping because of the Doppler broadening. The same applies to NHTX, 
where, however, only the first harmonic is simulated since the second is 
only marginally overdense and the OXB conversion region occurs in the core 
plasma rather than at the plasma edge. Moreover, the high second harmonic 
frequency ($\sim $100 GHz) combined with the relatively long density scale 
length in the conversion regions makes the OXB mode conversion angular 
window rather narrow.

\subsection{EBW system parameters}
\label{subsec:mylabel3}
As already shown in numerous previous works (see, e.g., \cite{42, 18}), the 
propagation path and the $N_\parallel $ evolution of EBW rays strongly 
depends on the vertical launch position. $N_\parallel $ appears in the 
resonance condition (\ref{eq12}) and can thus influence the wave absorption 
location. Moreover, $N_\parallel $ is an important factor in the EBW current 
drive. The vertical launch position is therefore a crucial parameter of any 
EBW launcher, which, besides the frequency, can be chosen arbitrarily. The 
toroidal and poloidal angles must be optimized for the conversion efficiency 
and we can only select negative or positive initial $N_{\parallel 0} $. 
These properties and restrictions dictate the EBW launcher parameters to be 
scanned in our survey: wave frequency, vertical launch position, and the 
sign of $N_{\parallel 0} $. The waves propagate far enough from the plasma 
top and bottom so that we can assume up-down symmetry. It has been verified 
for several cases that the above-midplane launcher power and current density 
profiles are symmetric with the below-midplane launcher profiles with the 
opposite sign of $N_{\parallel 0} $. In particular,
\begin{equation}
\label{eq19}
\begin{array}{rcr}
P\left( {\rho ,N_{\parallel 0} ,z_\mathrm{A} } \right) & = & P\left( {\rho 
,-N_{\parallel 0} ,-z_\mathrm{A} } \right), \\
 j\left( {\rho ,N_{\parallel 
0} ,z_\mathrm{A} } \right) & = & -j\left( {\rho ,-N_{\parallel 0} ,-z_\mathrm{A} } 
\right).
\end{array}
\end{equation}
The reason for this is that the flux surfaces are axially symmetric
with respect to the midplane.
As a result, EBW rays below and above the midplane are driven in 
opposite vertical directions with $N_\parallel $ having opposite signs. 
Subsequently, the ray paths are symmetric when starting from $\pm z_\mathrm{A} 
$ with opposite $N_\parallel $, and thus resulting in symmetric power 
deposition profiles. The opposite sign of $N_\parallel $ then causes a 
reversal in the driven current direction.

\begin{figure}[htbp]
\centering
\includegraphics[width=8cm]{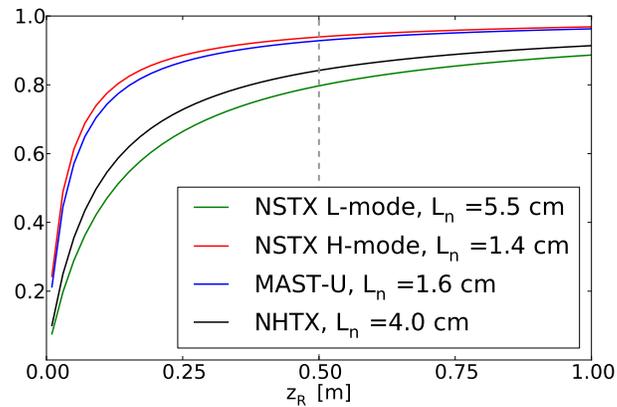}
\caption{Gaussian beam maximum conversion efficiency, \Eref{eq7} dependence on $z_\mathrm{R} $. The average $L_n $ in the mode conversion region is used. $\omega _{\mathrm{ce}} /\omega =0.5$.}
\label{fig2a}
\end{figure}

\begin{figure}[htbp]
\centering
\includegraphics[width=8cm]{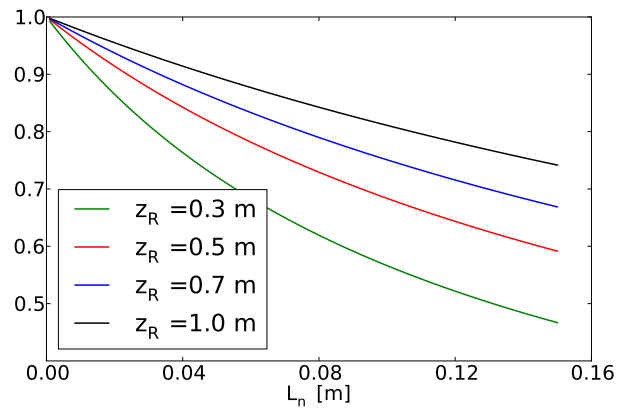}
\caption{Gaussian beam maximum conversion efficiency, \Eref{eq7} dependence on $L_n $. $\omega _{\mathrm{ce}} /\omega =0.5$.}
\label{fig2b}
\end{figure}

The frequencies and vertical launch positions used in our survey are given 
in Table \tablename~\ref{tab2}. The antenna beam Rayleigh range for 
all simulations is set to 0.5 m. In Figures \ref{fig2a} and \ref{fig2b}
are shown the maximum Gaussian beam O-X conversion efficiencies calculated 
from \Eref{eq7} for typical target plasma density scale lengths and the 
dependence on $L_n $ for various Rayleigh ranges. The selected $z_\mathrm{R} $ 
of 0.5~m is a compromise between small beams with poor conversion efficiency 
and large beams with high conversion efficiency but presumably wide power 
deposition profiles.

\begin{table}[htbp]
\caption{\label{tab2}EBW launcher system parameters used in this study.}
\small
\begin{tabular}{@{}l*{3}{p{3.9cm}}}
\br
scenario& 
1$^\mathrm{st}$ harmonic frequencies [GHz]& 
2$^\mathrm{nd}$ harmonic frequencies [GHz]& 
vertical launch positions [m] \\
\mr
NSTX L-mode& 
11, 11.5, 12, 12.5, 13, 14, 15, 16, 17, 18& 
21, 22, 23, 24, 25, 26, 27, 28, 29, 30& 
0, 0.1, 0.2, 0.3, 0.4, 0.5 \\
NSTX H-mode& 
12, 12.5, 13, 13.5, 14, 14.5, 15, 16, 17, 18, 19, 20& 
22, 23, 24, 25, 26, 27, 28, 29, 30& 
0, 0.1, 0.2, 0.3, 0.4, 0.5 \\
MAST-U& 
19, 20, 21, 22, 23, 24, 25, 26, 28& 
32, 33, 34, 35, 36, 37, 38, 39, 40, 41& 
0, 0.1, 0.2, 0.3, 0.4, 0.5, 0.6 \\
NHTX& 
52, 54, 56, 58, 60, 64, 68, 72, 76& 
none& 
0, 0.2, 0.4, 0.6, 0.8, 1.0, 1.2 \\
\br
\end{tabular}
\end{table}

\begin{figure}[htbp]
\centering
\includegraphics[width=0.9\linewidth]{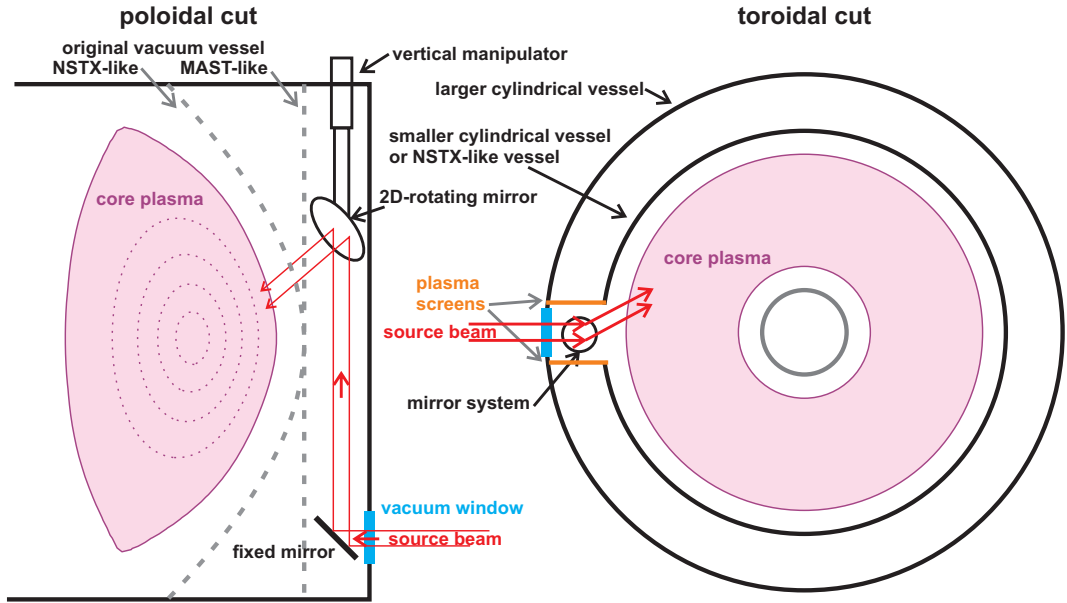}
\caption{A possible concept of an EBW launcher system with a 
vertically movable antenna. The design is sketched for two (existing) vacuum 
vessel shapes---NSTX-like (spherical) and MAST-like (cylindrical). Two 
variants are proposed---either inside a larger vessel or outside a smaller 
vessel.}
\label{fig3}
\end{figure}

A launcher that would span the whole parameter range listed in 
\tablename~\ref{tab2} is certainly unrealistic. Multi-frequency 
systems are rather challenging---even though such systems are actively being
studied \cite{43}. For these reasons we focus more on single-frequency systems. A 
concept of such a system is sketched in \figref{fig3}.
This system is designed to have a vertically movable mirror, which can 
be rotated in two dimensions (toroidally and poloidally), thus providing 
variable vertical launch positions with optimum OXB launch angles at the 
same time. In-vessel components are shielded from the plasma by screens that 
can either be part of the machine vacuum vessel (the smaller vessel variant) 
or placed inside a large MAST-like (cylinder) shape vessel (the larger 
vessel variant). This system is feasible with present day technologies and 
provides enough flexibility required for an advanced EBW system, as will be 
shown hereinafter. 

\begin{figure}[tbhp]
\begin{minipage}[b]{0.33\linewidth}
\centering
\includegraphics[width=0.95\linewidth]{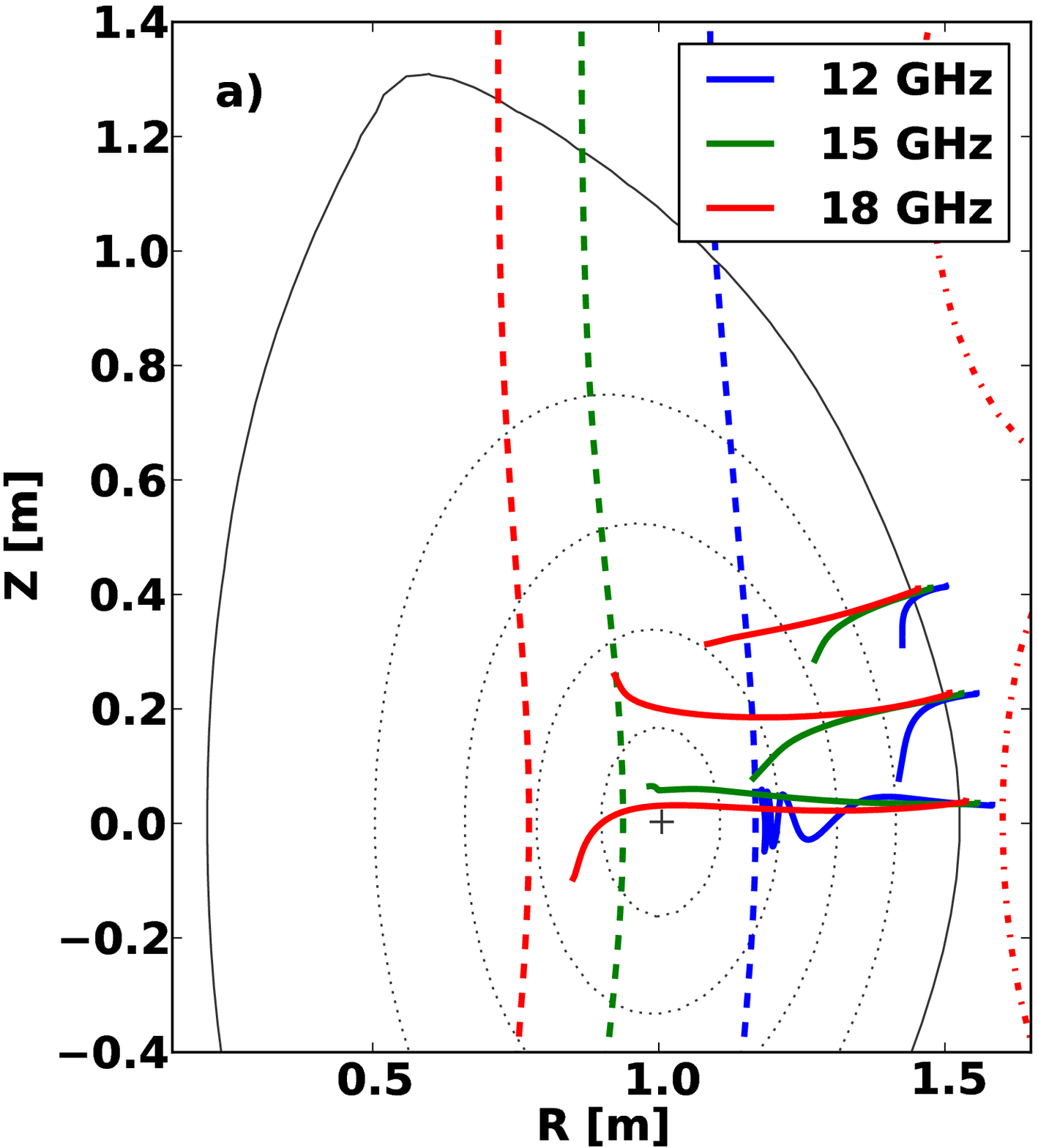}
\end{minipage}
\begin{minipage}[b]{0.33\linewidth}
\centering
\includegraphics[width=0.95\linewidth]{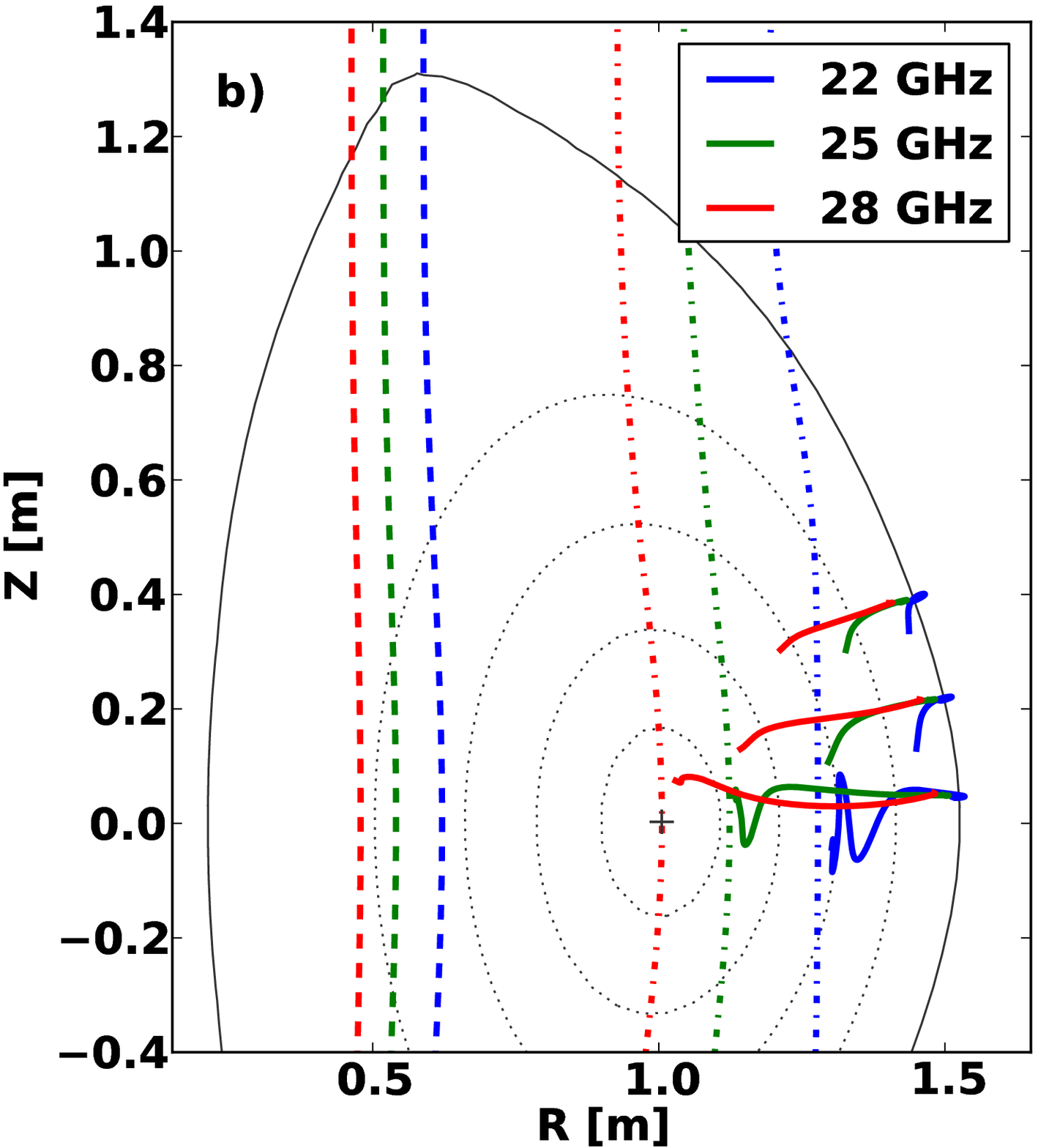}
\end{minipage}
\begin{minipage}[b]{0.33\linewidth}
\centering
\includegraphics[width=0.95\linewidth]{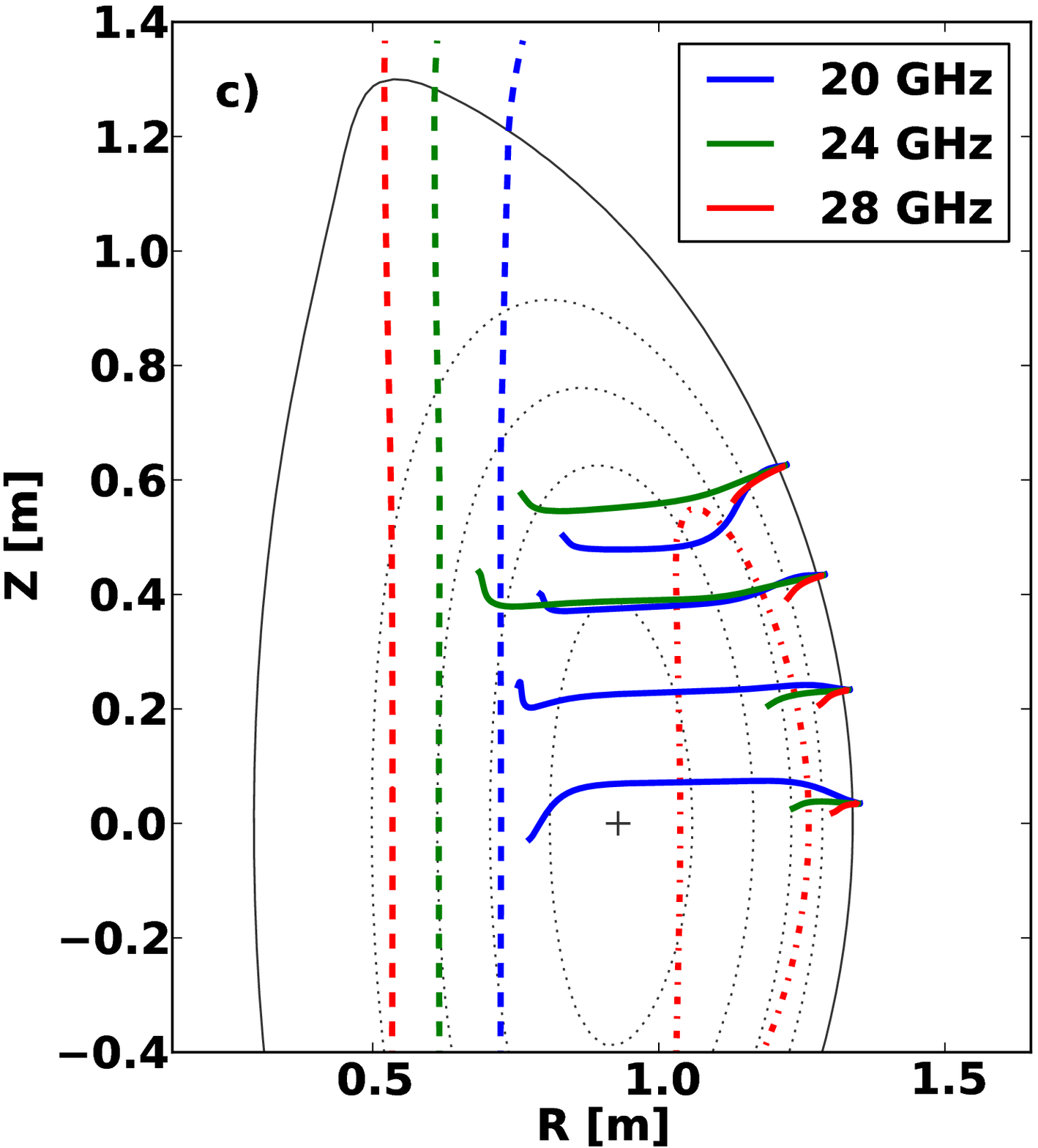}
\end{minipage}
\caption{Ray trajectories for various frequencies and vertical launch positions for a) NSTX L-mode, 1$^\mathrm{st}$ harmonic, b) NSTX L-mode 2$^\mathrm{nd}$ harmonic, c) MAST-U 1$^\mathrm{st}$ harmonic. Ray trajectories are plotted with solid lines, dashed and dash-dot lines show 1$^\mathrm{st}$ and 2$^\mathrm{nd}$ cold EC resonance surfaces, respectively.}
\label{fig4}
\end{figure}

The principle of using EBW control based on a vertically adjustable launcher 
is demonstrated in \figref{fig4}. It is clearly 
seen that the vertical launch position (besides the frequency) has a major 
effect on the ray propagation and strongly influences the location of the 
power deposition. We also notice the typical behaviour of EBW rays: Midplane 
rays with frequencies for which the cold EC resonance surfaces (typically 
concave shaped) occur in the inboard half of the plasma typically propagate 
straight until they reach the vicinity of a cold EC resonance, where their 
$\left| {N_\parallel } \right|$ grows exponentially and the rays are damped 
\cite{33}. Rays with lower frequencies, whose cold EC resonance layers appear in 
the outboard half with typically convex shape, oscillate around the midplane 
and their $N_\parallel $ oscillate around zero \cite{33}. Rays launched 
off-midplane are 
characterized by steadily and monotonically varying $N_\parallel $.
This behaviour is shown graphically in Section~\ref{sec:NparQLeff}.
 This 
results in a significant Doppler shift of the EC frequency and hence these 
waves are absorbed quite far from the cold resonance. 

\section{EBW H{\&}CD performance---numerical results}
\label{sec:mylabel1}

In this section, we present our result on EBW H{\&}CD performance in the previously 
listed scenarios. The mode conversion, which is a well-separable problem,
is always assumed to be 100{\%}.
There are two fundamental reasons for this assumption. First,
we would like to focus on EBW propagation, absorption and current drive.
These aspects, as will be shown hereinafter, are complex enough 
even when neglecting the mode conversion.
Second, the theoretical description of the mode conversion process is
so complex that we can arrive to almost any results depending on
which theories and processes are considered. Moreover, the knowledge
of the edge plasma, where the mode conversion occurs, is very limited.

\subsection{Localization and current drive efficiency}
\label{subsec:localization}
The current drive efficiency can be expressed in various ways. Most 
straightforward and suitable from the experimental and engineering point of 
view is the absolute efficiency
\begin{equation}
\label{eq20}
\eta \equiv \frac{I_{\mathrm{RF}} }{P_\mathrm{0} },
\end{equation}
where $I_{\mathrm{RF}} $ is the total current driven by the RF waves and 
$P_\mathrm{0} $ is the total injected RF power. However, the current drive 
efficiency unavoidably depends on plasma parameters, particularly the 
collisionality, and hence a quantity that reflects this intrinsic behaviour 
would be better suited for comparing different plasmas and current drive 
mechanisms. Commonly used for EC waves is the normalized efficiency $\zeta $ 
\cite{44}, which scales out the electron density and temperature collisional 
effects and the plasma size. 
However, $\zeta $ does not reflect the intrinsic effects of
particle trapping and effective ion charge ($Z_{\mathrm{eff}} )$.
The original definition assumes that 
the power deposition profile is well localized so the plasma parameters 
(density and temperature) do not change there. This is not always valid and 
we therefore use an absorbed power weighted average:
\begin{equation}
\label{eq21}
\zeta \equiv \frac{e^3}{\varepsilon _0^2 }\frac{R_0 }{P_\mathrm{0} }\int 
{\frac{n_\mathrm{e} \left( \rho \right)}{T_\mathrm{e} \left( \rho 
\right)}\frac{\mathrm{d}I_{\mathrm{RF}} }{\mathrm{d}\rho}\mathrm{d}\rho} .
\end{equation}
Here, $\mathrm{d}I_{\mathrm{RF}} $ is the RF driven current 
in a plasma surface enclosed by $\rho $ and 
$\rho +\mathrm{d}\rho $ flux surfaces. In the numerical simulations, LUKE 
selects a $\rho $-grid (based on the power deposition profile) so the 
volumes become finite:
\begin{equation}
\label{eq22}
\Upsilon _i \equiv V\left( {\frac{\rho _i -\rho _{i-1} }{2},\frac{\rho 
_{i+1} -\rho _i }{2}} \right),
\end{equation}
where $V\left( {\rho ,{\rho }'} \right)$ denotes the plasma volume enclosed 
by the flux surfaces $\rho $ and ${\rho }'$. $\rho _i $ are the LUKE grid 
points. The discrete form of (\ref{eq21}) is then
\begin{equation}
\label{eq23}
\zeta =\frac{e^3}{\varepsilon _0^2 }\frac{R_0 }{P_\mathrm{0} }\sum\limits_i 
{\frac{n_\mathrm{e} \left( {\rho _i } \right)}{T_\mathrm{e} \left( {\rho _i } 
\right)}I_i } \cong 3.27\frac{R_0 \left[ \mathrm{m} \right]}{P_\mathrm{0} \left[ 
\mathrm{W} \right]}\sum\limits_i {\frac{n_\mathrm{e} \left( {\rho _i } 
\right)\left[ {\mathrm{10}^{19}\mathrm{m}^{\mathrm{-3}}} \right]}{T_\mathrm{e} 
\left( {\rho _i } \right)\left[ {\mathrm{keV}} \right]}I_i \left[ \mathrm{A} 
\right]} ,
\end{equation}
where $I_i $ is the current driven in the poloidal cross-section of 
$\Upsilon _i $. Note that $\zeta $ reflects the sign of the driven current.

\begin{figure}[htbp]
\centering
\includegraphics[width=0.9\linewidth]{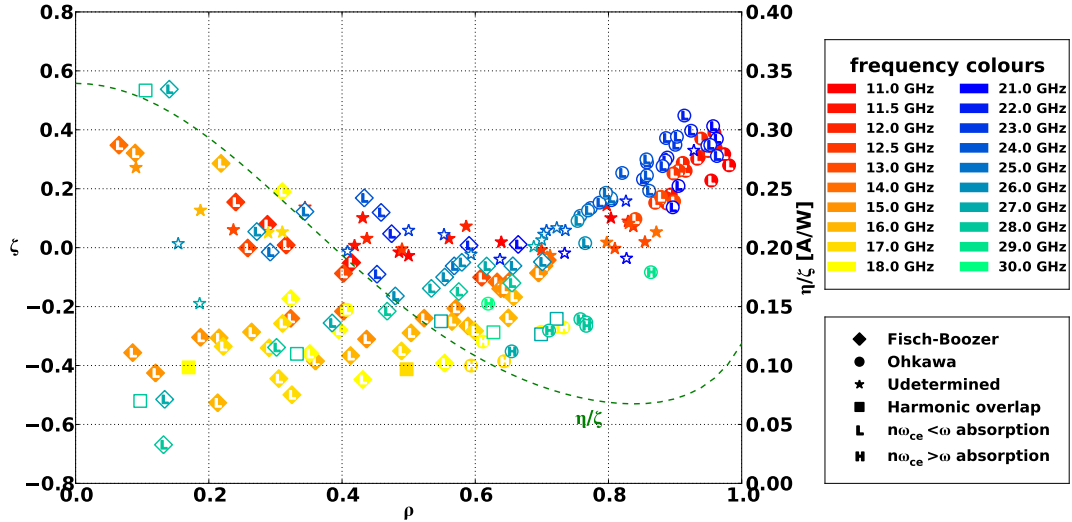}
\caption{Current drive efficiency $\zeta $(symbols) 
and $\eta /\zeta $ conversion factor (dashed line) versus $\rho 
$, NSTX L-mode first (full symbols) and second (open symbols) 
harmonics, all frequencies and vertical launch positions as listed in 
\tablename~\ref{tab2}, both positive and 
negative $N_{\parallel 0} $, 1 MW incident power. Neither the 
vertical launch position nor the $N_{\parallel 0} $ sign can be 
graphically distinguished in the figure.}
\label{fig5}
\end{figure}

\begin{figure}[htbp]
\centering
\includegraphics[width=0.9\linewidth]{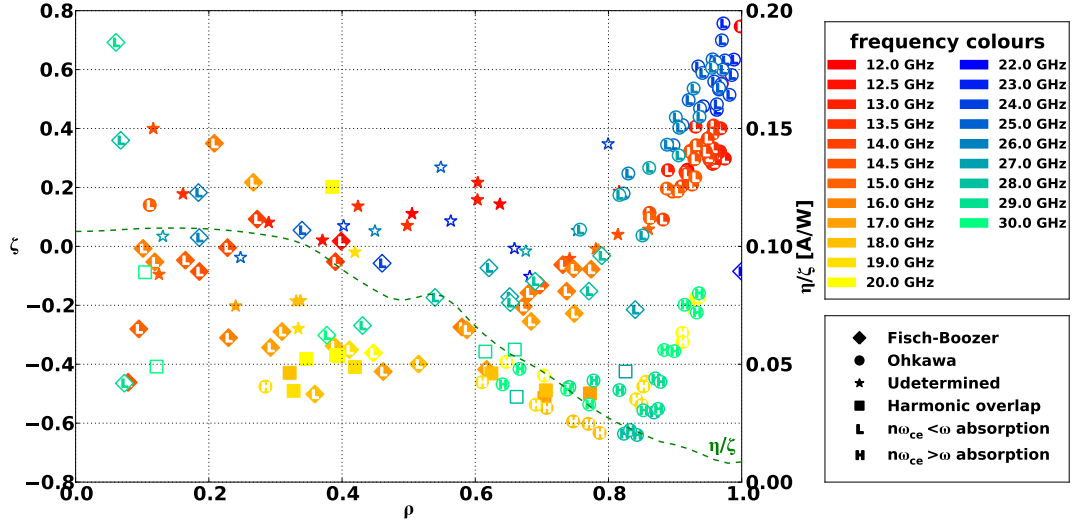}
\caption{Same as \figref{fig5} -- but for the NSTX H-mode.}
\label{fig6}
\end{figure}

\begin{figure}[htbp]
\centering
\includegraphics[width=0.9\linewidth]{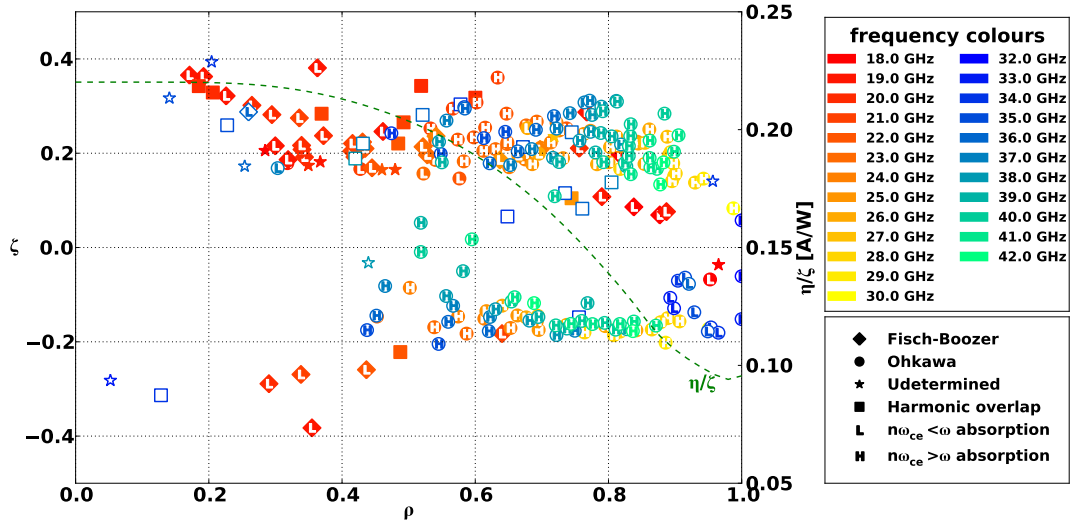}
\caption{Same as \figref{fig5} -- but for the MAST-U plasma.}
\label{fig7}
\end{figure}

\begin{figure}[htbp]
\centering
\includegraphics[width=0.9\linewidth]{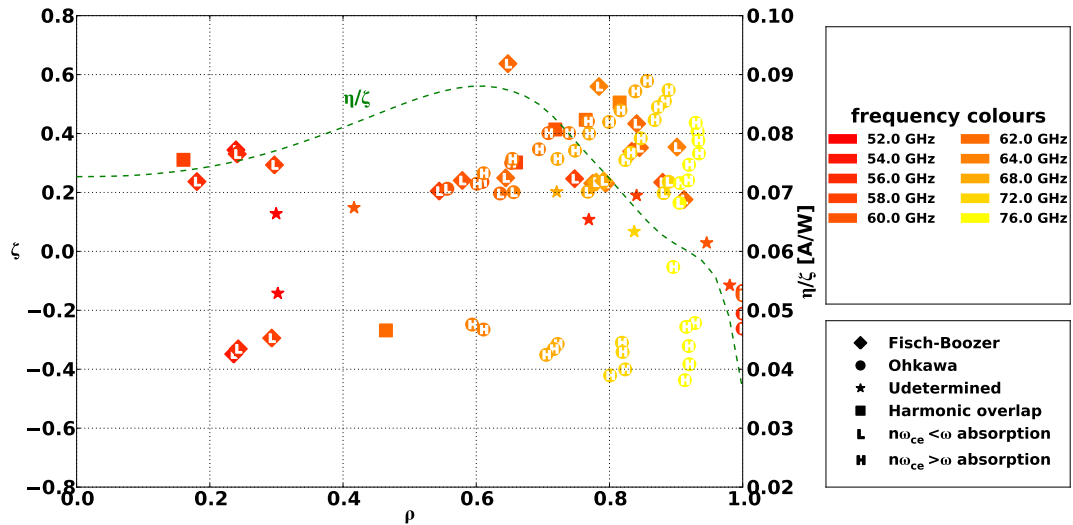}
\caption{Same as \figref{fig5} -- but for the NHTX plasma.}
\label{fig8}
\end{figure}

In Figures \ref{fig5} -- \ref{fig8} we show the current drive efficiency 
$\zeta $ for all the plasma and launch scenarios listed in 
\tablename~\ref{tab2}, i.e., for the different frequencies, 
vertical launch positions and toroidal injection directions. 
The classification of the current drive mechanism is performed automatically
by calculating the average (absorbed power weighted) $N_\parallel$ of the rays
and subsequently comparing the LUKE-calculated current direction to
Ohkawa and Fisch-Boozer current directions. In certain cases, this leads to ambiguous
results, either because the rays have mixed signs of $N_\parallel$ or they are
absorbed at different harmonics.
The results 
were obtained by AMR and LUKE coupled simulations with 1 MW incident power. 
We immediately notice the importance of the launch parameters as they strongly 
influence the location of the wave power deposition (which obviously 
coincides with the driven current location) and the current drive efficiency 
in a fixed plasma equilibrium. Clearly, by changing these parameters, we can 
select a specific scenario---on/off axis deposition at almost any $\rho $ 
with high/low $\left| \zeta \right|$. There is full flexibility in the 
direction of the driven current because of the (a)symmetry (\ref{eq19}). EBWs are 
most flexible and efficient in driving current in NSTX plasmas, mainly 
because in this case the magnetic field is monotonic without any magnetic well 
in the edge region. $\left| \zeta \right|\cong 0.4$ can be reached at almost 
any radius in NSTX. Our current drive efficiencies are similar to 
experimental values from COMPASS-D \cite{6} or Wendelstein 7-AS \cite{8} as well as to 
numerical results obtained for MAST-U \cite{45} or NSTX \cite{46}.

While similar efficiencies can be reached using EC X- and O-modes in the 
central region, the X- and O-mode current drive efficiency typically 
decreases with radius, 
particularly because of trapping effects (see, e.g., \cite{44}), which is not the case with EBWs. 
The L-mode plasma parameters cause higher absolute current drive efficiency, 
i.e., higher $\eta /\zeta $. There exist several significantly higher 
efficiency second harmonic cases with $\rho \cong 0.1$ (i.e., almost on the 
magnetic axis) in both L- and H-modes.
Even though the results reported here indicate a limited on-axis 
accessibility and flexibility, this may not necessarily hold for slightly different 
frequencies or launch positions.

Typically we find that in the central plasma regions we drive a Fisch-Boozer 
current \cite{47} while Ohkawa current \cite{48} in edge regions. Higher harmonic 
absorption, i.e., absorption on the $n^{\mathrm{th}}$ harmonic with $n\omega 
_{\mathrm{ce}} >\omega $, favours the Ohkawa mechanism. Typically we can 
distinguish three EBW efficient current drive regions:

\begin{enumerate}
\item Fisch-Boozer current drive with lower harmonic absorption predominantly near the centre.
\item Ohkawa current drive with lower harmonic absorption predominantly near the edge.
\item Ohkawa current drive with higher harmonic absorption predominantly between the plasma centre and edge (i.e., between the first and second regions).
\end{enumerate}
The location and size of these regions are very different in the 
investigated plasmas and these regions typically overlap. There also exist 
cases when the harmonics are overlapping, i.e., the wave is absorbed on two 
different EC harmonics. Quite interestingly, in these cases the current is 
still driven in one direction even though the resonant electrons have their 
$v_{\parallel \mathrm{res}} $ having different signs. This occurs because the 
lower harmonic absorption favours the Fisch-Boozer mechanism (for which 
$v_{\parallel \mathrm{res}} \cdot j<0)$, while the higher harmonic absorption 
favours the Ohkawa mechanism (for which $v_{\parallel \mathrm{res}} \cdot 
j>0)$. 

In going from an NSTX L-mode to an NSTX H-mode to a MAST-U and then to an 
NHTX plasma, the external magnetic field increases together with the 
appearance of magnetic wells near the edge (these wells being caused by 
strong edge currents) and we observe a decrease in the current drive 
efficiency, as well a decrease in the flexibility of the EBW absorption and 
central plasma accessibility. 

\subsection{\label{sec:NparQLeff}$N_\parallel $ and quasilinear effects}

\begin{figure}[htbp]
\centering
\includegraphics[width=8cm]{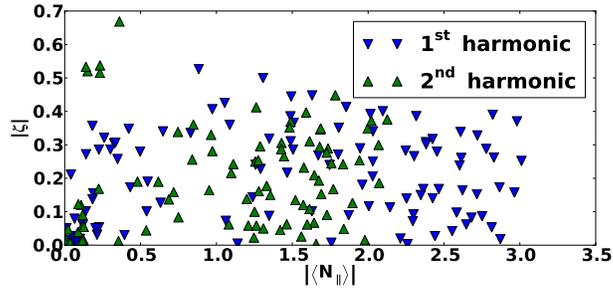}
\caption{Current drive efficiency versus the magnitude of the mean $N_\parallel$, for all NSTX L-mode cases, $P_0=1$ MW.}
\label{fig9}
\end{figure}

\begin{figure}[htbp]
\centering
\includegraphics[width=8cm]{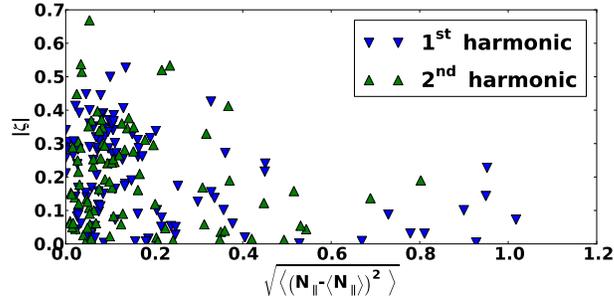}
\caption{Current drive efficiency versus absolute variance of $N_\parallel$, for all NSTX L-mode cases, $P_0=1$ MW.}
\label{fig10a}
\end{figure}

\begin{figure}[htbp]
\centering
\includegraphics[width=8cm]{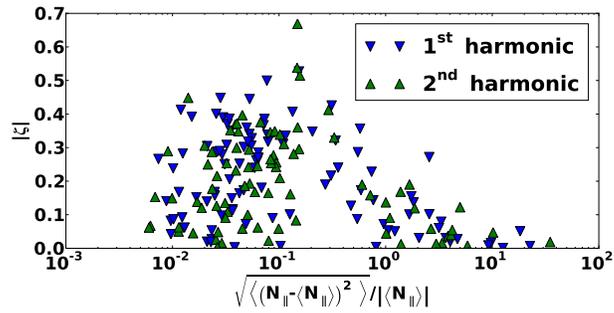}
\caption{Current drive efficiency versus relative variance of $N_\parallel$, for all NSTX L-mode cases, $P_0=1$ MW.}
\label{fig10b}
\end{figure}

In order to show the effects of $N_\parallel $, we calculate its average 
value (not to be confused with the initial $N_\parallel )$, weighted by the 
absorbed power:
\begin{equation}
\label{eq24}
\left\langle {N_\parallel } \right\rangle =\frac{\sum\limits_i {\Delta 
P\left( {\Upsilon _i } \right)\frac{\sum\limits_{\mathrm{rays,}\;\rho \in 
\Upsilon _i } {N_\parallel \Delta P_{\mathrm{ray}} \left( {N_\parallel ,\rho } 
\right)} }{\sum\limits_{\mathrm{rays,}\;\rho \in \Upsilon _i } {\Delta 
P_{\mathrm{ray}} \left( {N_\parallel ,\rho } \right)} }} }{\sum\limits_i 
{\Delta P\left( {\Upsilon _i } \right)} }.
\end{equation}
\figref{fig9} shows the current drive efficiency 
versus $\left| {\left\langle {N_\parallel } \right\rangle } \right|$. It is 
found that the two quantities are clearly uncorrelated. 
As a consequence of the short wavelength of EBWs ($k_\perp \rho_{\rm e} \sim 1$),
the resonant $v_\perp$ is low, irrespective of the value of $N_\parallel$.
The dominant factors in EBW CD efficiency are the $N_\parallel$-spectrum (mixing of signs), 
harmonic overlapping (because of large $\left| N_\parallel \right|$)
and Fisch-Boozer versus Ohkawa effects.
Figures \ref{fig10a} and \ref{fig10b} show the current drive efficiency 
versus the absolute and the relative $N_\parallel $ variance: again, we find 
no clear correlation. 
There is only a weak (logarithmic) decrease with the 
relative variance, starting at $\sim $0.1. Most of the cases have a rather 
narrow $N_\parallel $ spectrum, with absolute variance $<0.2$ and relative 
variance $<0.1$. NSTX H-mode, MAST-U and NHTX results show very similar 
behaviour.

\begin{figure}[htbp]
\centering
\includegraphics[width=0.9\linewidth]{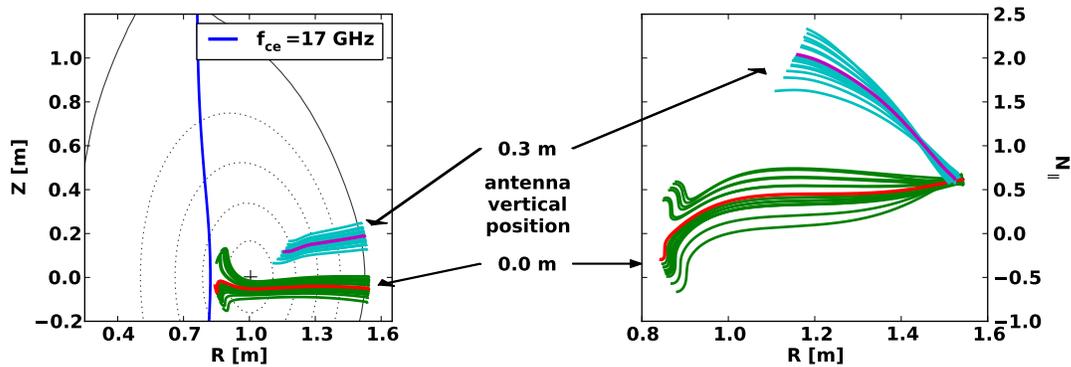}
\caption{Ray trajectories and the evolution of $N_\parallel $ for two NSTX L-mode cases at 17 GHz with different vertical launch positions: 0 m and 0.3 m.}
\label{fig11}
\end{figure}

We now compare the effect of two different vertical launch positions for the 
NSTX L-mode plasma at frequency 17 GHz. The ray trajectories and the 
evolution of $N_\parallel $ are plotted in 
\figref{fig11}. Those rays launched close to the 
midplane propagate straight to the magnetic axis, and the central ray's 
$N_\parallel $ does not change appreciably until the ray gets close to the 
resonance (around $R=1\;\mathrm{m})$. $\left| {N_\parallel } \right|$ now 
starts to increase exponentially, and the beam splits in two parts that 
propagate in opposite vertical directions. Finally, the rays are absorbed, 
having been split in approximately two halves with opposite signs of 
$N_\parallel $ at the absorption location. This behaviour demonstrates the 
typical behaviour of midplane rays at frequencies where cold EC resonance 
surface is lying in the inboard half of the plasma. For off-midplane launch 
at 17 GHz, one sees in \figref{fig11} that $\left| 
{N_\parallel } \right|$ steadily increases and the waves are absorbed at an 
EC resonance that has been Doppler shifted. Since all the rays have the same 
sign in $N_\parallel $ signs, one achieves a high current drive efficiency. 

\begin{figure}[htbp]
\centering
\includegraphics[width=8cm]{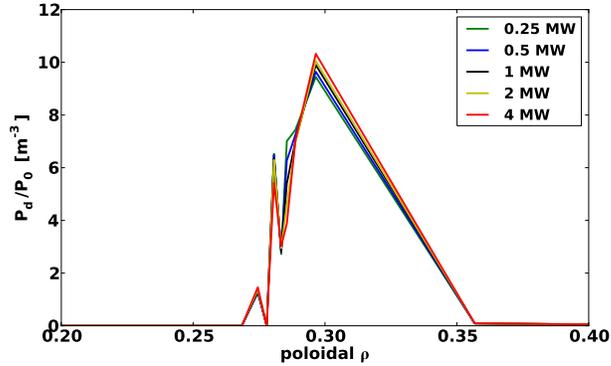}
\caption{Power deposition radial profile, NSTX L-mode, 17 GHz, 0.0 m vertical launch position.}
\label{fig12a}
\end{figure}

\begin{figure}[htbp]
\centering
\includegraphics[width=8cm]{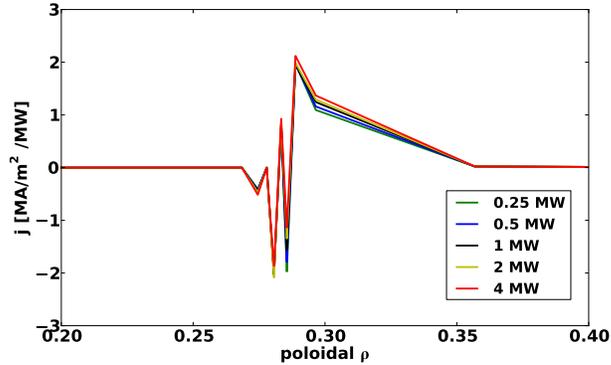}
\caption{Driven current density radial profile, NSTX L-mode, 17 GHz, 0.0 m vertical launch position.}
\label{fig12b}
\end{figure}

\begin{figure}[htbp]
\centering
\includegraphics[width=8cm]{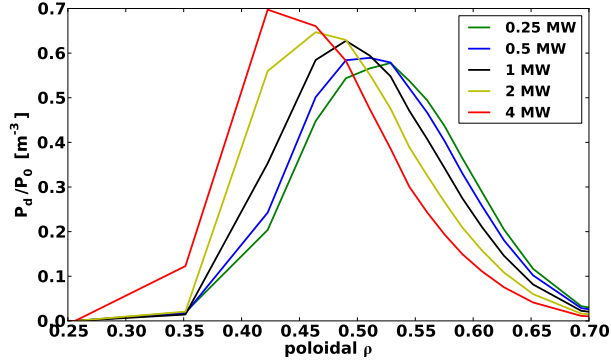}
\caption{Power deposition radial profile, NSTX L-mode, 17 GHz, 0.3 m vertical launch position.}
\label{fig13a}
\end{figure}

\begin{figure}[htbp]
\centering
\includegraphics[width=8cm]{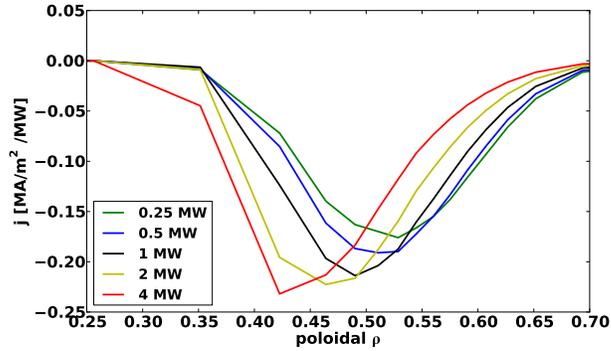}
\caption{Driven current density radial profile, NSTX L-mode, 17 GHz, 0.3 m vertical launch position.}
\label{fig13b}
\end{figure}

These two examples clearly demonstrate how the deposition location and the 
current drive efficiency can be controlled by the choice of the vertical 
launch position. In Figures \ref{fig12a} -- \ref{fig13b} we see the resulting power 
deposition and driven current densities, plotted for launched powers from 
0.25 MW to 4 MW. The power deposition profile (and consequently the driven 
current profile) is rather narrow in the midplane launch case 
(Figures \ref{fig12a} and \ref{fig12b}), since there is a sharp resonance 
close to the cold EC resonance surface. The driven current profile is 
oscillating around zero, resulting in nearly zero net driven current.
These oscillations are rather artificial, partly because of the beam 
discretization by individual rays, and partly because,
in reality, such oscillations would most probably be smoothed out by radial 
transport.
There is almost no dependence on the launched power because of 
the sharp resonance.

However, in the 0.3 m vertical launch position case 
(Figures \ref{fig13a} and \ref{fig13b}), both the power deposition profile 
and the current drive profile are much broader. This is due to the strongly 
Doppler-shifted absorption with a relatively large $N_\parallel $ spectral 
width, as well as Doppler broadening effects \cite{31}. The current is driven in 
one direction as the sign of $N_\parallel $ is identical for all the rays. 
For this case there is power deposition on overlapping EC harmonics. 
Moreover, this is one of the interesting cases mentioned in the previous 
section, in which the Fisch-Boozer current from the deposition on the lower 
harmonic and the Ohkawa current from the deposition on the higher harmonic 
are in the same direction.

In this configuration, the power is deposited on suprathermal electrons \cite{31} 
and EBW H{\&}CD is therefore strongly affected by quasilinear effects. 
Quasilinear flattening of the distribution function with increasing power 
levels yields a relative reduction of the absorbed power, resulting in a 
more inward deposition.

\begin{figure}[htbp]
\centering
\includegraphics[width=8cm]{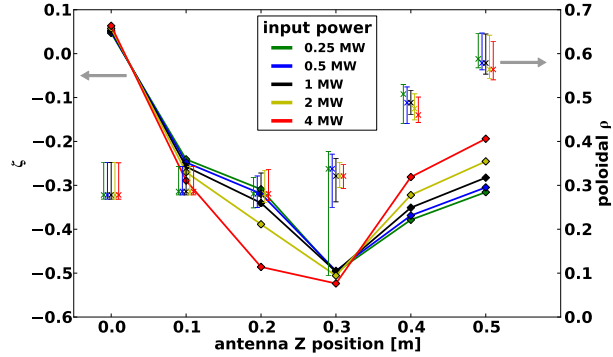}
\caption{Input power scan of current drive efficiency and radial location of the current peak for 17 GHz NSTX L-mode cases with positive initial  $N_\parallel $. Line-plots with symbols (colour online) represent $\zeta $ while symbols with vertical error bars (colour online) represent the radial current location $\rho _{j\max } $ and its width $\sigma _j $ (upper and lower limits represent $\rho _{j\pm 1} $). Each group of the cross symbols belongs to one antenna position, and a horizontal shift is employed to separate the lines visually.}
\label{fig14a}
\end{figure}

\begin{figure}[htbp]
\centering
\includegraphics[width=8cm]{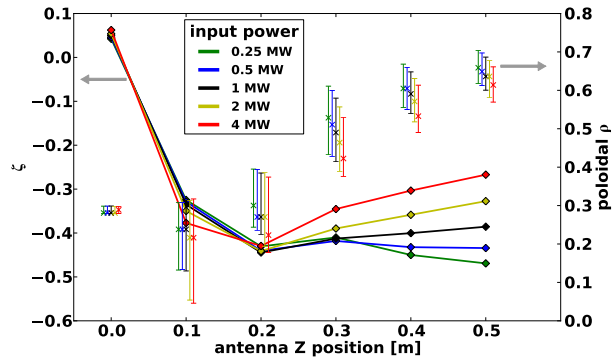}
\caption{Same as \figref{fig14a}, but for negative initial $N_\parallel $.}
\label{fig14b}
\end{figure}

An input power scan for several NSTX L-mode cases is presented in 
Figures \ref{fig14a} and \ref{fig14b}.  We find that there is no general 
tendency of the current drive efficiency to increase or decrease with the 
input power. Cases exist with increasing, decreasing or invariant $\zeta $ 
dependence on input power. However, in most cases, increasing power leads to 
either lower or similar current drive efficiency. In 
Figures \ref{fig14a} and \ref{fig14b} we also show the current profile 
maximum radial location $\rho _j $ and its width $\sigma _j $, defined as
\begin{equation}
\label{eq25}
\rho _{j\max } \equiv \mathop {\arg \max }\limits_{\rho \in \left[ {0,1} 
\right]} \left| {j\left( \rho \right)} \right|,
\end{equation}
\begin{equation}
\label{eq26}
\rho _{j-1/2} \equiv \min \left\{ {\rho :\left| {j\left( \rho \right)} 
\right|=\left| {j\left( {\rho _{j\max } } \right)/2} \right|\wedge 
0\leqslant \rho <\rho _{j\mathrm{max}} } \right\},
\end{equation}
\begin{equation}
\label{eq27}
\rho _{j+1/2} \equiv \max \left\{ {\rho :\left| {j\left( \rho \right)} 
\right|=\left| {j\left( {\rho _{j\max } } \right)/2} \right|\wedge \rho 
_{j\mathrm{max}} <\rho \leqslant 1} \right\},
\end{equation}
\begin{equation}
\label{eq28}
\sigma _j \equiv \rho _{j+1/2} -\rho _{j-1/2} .
\end{equation}
In other words, $\sigma _j $ corresponds to the full width at half maximum 
if the current profile is considered single-peaked. Increasing power causes 
the wave absorption to occur further along the direction of propagation, which 
can either be towards the axis if the absorption occurs on the outboard side 
or away from the axis in the opposite case. This is caused by the 
quasilinear flattening of the distribution function and consequently lower 
absorption rate. 

\subsection{The effect of $Z_{\mathrm{eff}} $}
\label{subsec:mylabel4}
\begin{figure}[htbp]
\centering
\includegraphics[width=8cm]{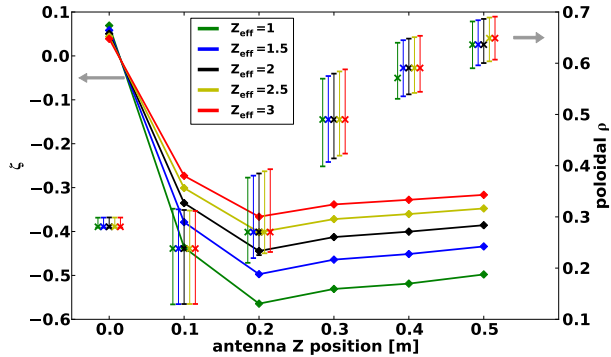}
\caption{$Z_{\mathrm{eff}} $ scan of current drive efficiency and radial location of the current peak for 17 GHz NSTX L-mode cases with negative initial $N_\parallel $, $P_0=1$ MW. Line-plots with symbols (colour online) represent $\zeta $ while symbols with vertical error bars (colour online) represent the radial current location $\rho _{j\max } $ and its width $\sigma _j $ (similarly to \figref{fig14a}).}
\label{fig15a}
\end{figure}

\begin{figure}[htbp]
\centering
\includegraphics[width=8cm]{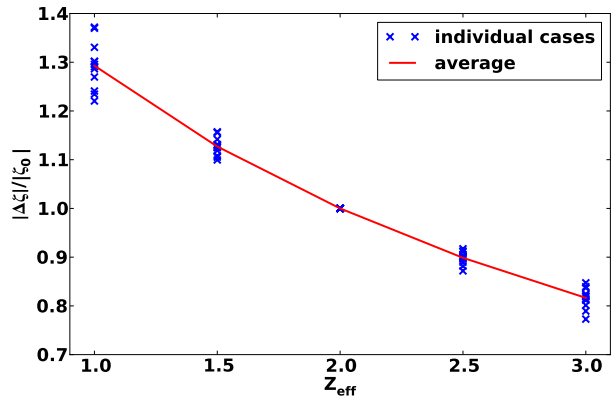}
\caption{Relative (with respect to $Z_{\mathrm{eff}} =2$ values, denoted $\zeta _0 $) changes of the current drive efficiency versus $Z_{\mathrm{eff}} $ for all 17 GHz NSTX L-mode cases with both positive and negative initial $N_\parallel $, $P_0=1$ MW.}
\label{fig15b}
\end{figure}

So far we assumed $Z_{\mathrm{eff}} =2$, which is a realistic experimental 
value. However, $Z_{\mathrm{eff}} $ can vary and it is very important to show 
the effect on EBW performance. We employ here the cases from the previous 
section and rerun the simulations with $Z_{\mathrm{eff}} $ ranging from 1 to 
3. \figref{fig15a} shows the effect of 
$Z_{\mathrm{eff}} $ on the current drive efficiency and the position of the 
current peak. $Z_{\mathrm{eff}} $ affects the electron-ion collision frequency 
and particularly pitch-angle scattering. A larger value of $Z_{\mathrm{eff}} $ 
results in faster isotropization of current-carrying fast electrons. Thus, 
the current drive efficiency is inversely proportional to $Z_{\mathrm{eff}} $ 
\cite{47}. The general trend of $\zeta $ versus $Z_{\mathrm{eff}} $ is shown in 
\figref{fig15b}. Compared to $Z_{\mathrm{eff}} =2$ 
results, the EBW current drive efficiency increases on average by 29 {\%} 
for $Z_{\mathrm{eff}} =1$ while a decrease of 18 {\%} is observed for 
$Z_{\mathrm{eff}} =3$. There is also a minor effect of $Z_{\mathrm{eff}} $ on 
the EBW deposition location, as can be seen in 
\figref{fig15a}. This is again caused by the 
collision frequency change, which affects the plasma quasi-linear response 
to the wave power.

\subsection{Robustness}
\label{subsec:robustness}
An important factor of any H{\&}CD system is its robustness---the 
sensitivity to changes in plasma conditions or in the system's parameters. 
In the previous section we have already investigated what happens if the 
injected power is changed. Moreover, the effect of changing the vertical 
launch position can be seen in Figures \ref{fig14a} and \ref{fig14b}; this 
effect is rather strong and therefore the vertical launch position must be 
carefully chosen and controlled. In this section we focus on the EBW H{\&}CD 
performance sensitivity to plasma parameters.

\begin{figure}[htbp]
\centering
\includegraphics[width=8cm]{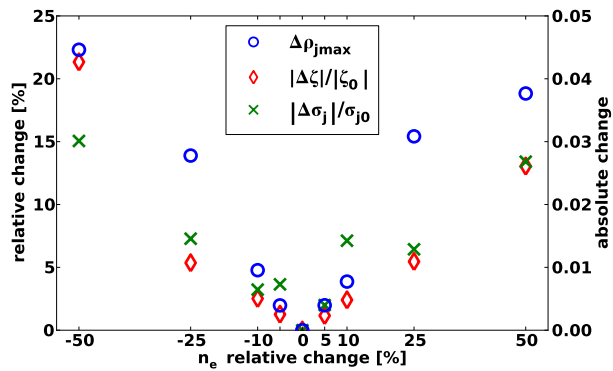}
\caption{Medians of absolute current location difference $\Delta \rho _{j\max } $ and relative current drive efficiency and current profile width differences versus varying plasma electron density. 17 GHz NSTX L-mode 1 MW cases (similarly to \figref{fig14a}) are used to calculate the medians.}
\label{fig16a}
\end{figure}

\begin{figure}[htbp]
\centering
\includegraphics[width=8cm]{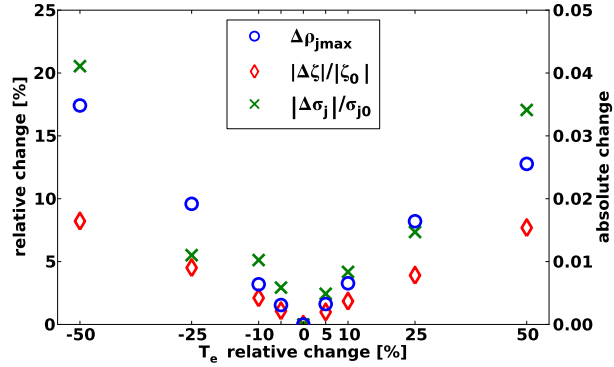}
\caption{Same as \figref{fig16a}, but for varying electron temperature}
\label{fig16b}
\end{figure}

In Figures \ref{fig16a} and \ref{fig16b} we show the sensitivity of EBWs 
to plasma electron density and temperature variations in $\pm $50 {\%} 
range. All vertical launch positions and both initial $N_\parallel $ signs 
of 17 GHz NSTX L-mode cases are used to calculate the medians of absolute 
location difference $\Delta \rho _{j\max } $ and relative current drive 
efficiency and profile widths $\left| {\Delta \zeta } \right|/\left| {\zeta 
_0 } \right|$ and $\left| {\Delta \sigma _j } \right|/\left| {\sigma _{j0} } 
\right|$, where the 0 subscripts denote results with the original plasma 
profiles. We first see a monotonic dependence of all the plotted quantities 
(except for two cases in $\left| {\Delta \sigma _j } \right|)$, indicating a 
non-chaotic behaviour of EBW performance with changing plasma profiles. 
Quantitatively, the radial current location changes fractionally compared to 
the typical $\sigma _j \sim 0.1$. However, very precise localization might 
be important for certain applications, in which case a feedback system is 
highly advisable. The median difference in current drive efficiency is below 
5 {\%} for less than 25 {\%} changes in the plasma profiles, which is very 
favourable. The current profile width is slightly more sensitive, a 
consequence of Doppler broadening. Not shown here are the variances. 
However, highest sensitivity is generally observed at lower frequencies, 
close to a midplane launch where the rays tend to oscillate, leading to 
current drive efficiencies that are typically low. In most cases the 
results are close to the median values.

\begin{figure}[htbp]
\centering
\includegraphics[width=8cm]{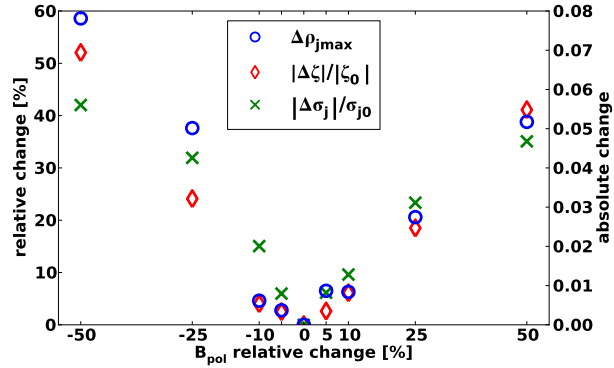}
\caption{Same as Figure \ref{fig17a}, but for varying poloidal magnetic field.}
\label{fig17a}
\end{figure}

\begin{figure}[htbp]
\centering
\includegraphics[width=8cm]{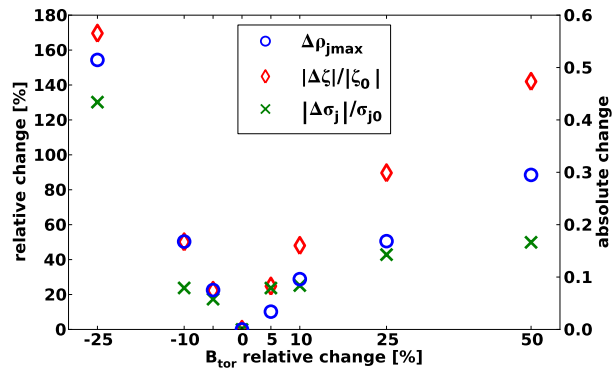}
\caption{Same as Figure \ref{fig17a}, but for varying toroidal magnetic field. Results for $B_{\mathrm{tor}} $ -50 {\%} could not be calculated.}
\label{fig17b}
\end{figure}

Another parameter that can vary in tokamaks is the plasma current and the 
toroidal magnetic field. Unlike density and temperature profiles, which are 
only crudely pre-programmed and evolve during the discharge, it is typical 
that the plasma current and toroidal magnetic field do not change during the 
discharge (except, of course, in the start-up and shut-down phases) and that 
their properties are pre-programmed with high confidence. This makes the 
demands on the sensitivity on these quantities less stringent as compared to 
the temperature and the density. In Figures \ref{fig17a} and \ref{fig17b} 
we show the sensitivity of 17 GHz L-mode cases to poloidal and toroidal 
magnetic field changes. The fields are simply changed by multiplying the 
respective components so that the resulting equilibrium is no longer a 
solution of the Grad-Shafranov equation. Significantly larger effects of the 
magnetic field changes on EBW results can immediately be noticed. The 
sensitivity is particularly high for the toroidal field simply because the 
toroidal field is much larger than the poloidal field in most of the plasma 
cross-section. Also notice that changing the total magnetic field by 10 {\%} 
is similar to changing the heating frequency by 1.4 GHz, which is the change 
in the central $\omega _{\mathrm{ce}} $. For large magnetic field changes we 
can even change the EC absorption harmonic number---e.g., decreasing 
$B_{\mathrm{tor}} $ by 25 {\%} shifts 17 GHz into the second harmonic range.

\section{Conclusions}
\label{sec:conclusions}
By means of coupled ray-tracing and Fokker-Planck simulations, we have 
thoroughly investigated electron Bernstein wave H{\&}CD prospects for 
spherical tokamaks. For the first time, a simple analytic formula for the 
O-X conversion efficiency of a Gaussian beam is derived from 1D plane wave 
theory. This formula supports our choice of the Rayleigh range as the 
antenna beam principal parameter that is fixed for all simulated cases. On 
an extensive set of EBW launch scenarios with varying frequency, vertical 
antenna position and toroidal injection angle, we show that EBWs can be 
absorbed at almost arbitrary radius and that EBWs can drive current with 
efficiencies comparable to electron cyclotron O- or X-modes. Moreover, the 
efficiency does not change with radius, while typically the efficiency of X- 
and O-modes decreases with radius. Best results in terms of efficiency and 
flexibility are achieved in NSTX plasmas, where the electron cyclotron 
frequency radial profiles are monotonic. In general, normalized current 
drive efficiencies $\left| \zeta \right|$ on the order of 0.3 -- 0.4 are 
feasible for all target plasmas, absolute efficiencies then depend on the 
plasma parameters as $I_{\mathrm{RF}} /P_0 \cong 0.31\zeta T_\mathrm{e} /R_0 
n_\mathrm{e} $, where the units are keV for $T_\mathrm{e} $, m for $R_0 $ and 
$10^{19}\mathrm{m}^{-3}$ for $n_\mathrm{e} $. These results, however, do not
reflect the mode conversion efficiency, which, as we have shown, is limited
by the beam divergence and can be further degraded be other effects.

For EBWs, the initial value of $\vert N_\parallel \vert $ is fixed by the 
mode-conversion process and only the sign of $N_\parallel $ can be chosen at 
will. The evolution of $N_\parallel $ is determined by the wave frequency, 
the vertical launch position and by the plasma parameters. We have shown how 
different vertical launch positions strongly influence the $N_\parallel $ 
spectrum and consequently the current drive efficiency. However, there seems 
to be no general correlation between the current drive efficiency and the 
$N_\parallel $ spectrum and its width. 

Input power scans have been performed to investigate the quasilinear 
effects. Increasing power generally leads to either lower or similar current 
drive efficiency. Higher power also causes the wave absorption to occur 
further along the direction of propagation, which can either be towards the 
axis if the absorption occurs on the outboard side or away from the axis in 
the opposite case. An important factor is the effective ion charge, which 
affects the electron-ion collisionality, and, consequently, the current 
drive efficiency significantly depends on $Z_{\mathrm{eff}} $. A minor effect of 
$Z_{\mathrm{eff}} $ on the driven current location can be observed, which is 
caused by changing the plasma quasilinear response.

The sensitivity of EBW H{\&}CD to changes in plasma parameters has been 
investigated. It has been shown that the EBW performance is rather robust. 
Neither the current drive efficiency nor the radial location changes 
significantly when the electron temperature or density changes moderately. 
Larger sensitivity is observed for magnetic field changes, especially the 
(dominant) toroidal field.

In conclusion, the EBW is a promising candidate for a powerful and flexible 
auxiliary H{\&}CD system for spherical tokamaks, in many aspects comparable 
to EC systems for standard aspect ratio tokamaks. However, several problems, 
which are not including in this survey, particularly the coupling,
need to be addressed and better understood. High-power experiments on
major spherical tokamaks or similar machines would be greatly beneficial
to tackle these topics and validate our results.

\ack
This work was supported by the grant no. 202/08/0419 of Czech Science Foundation, the Academy of Sciences of the Czech Republic IRP {\#}AV0Z20430508, the Ministry of Education, Youth and Sports CR {\#}7G10072, U.S. Department of Energy DE-AC02-09CH11466 and European Communities under the contract of Association between EURATOM/IPP.CR No. FU07-CT-2007-00060 and under the contract of Association between Euratom and CEA and carried out within the framework of the European Fusion Development Agreement. The views and opinions expressed herein do not necessarily reflect those of the European Commission.

\end{document}